\documentclass[a4paper,10pt]{article}
\usepackage{amsfonts}
\usepackage{amsmath,amssymb}
\usepackage{graphicx}

\makeatletter \@addtoreset{equation}{section}
\def\Z{\mathbb Z}

\def\R{\mathbb R}
\def\be{\begin{equation}}
\def\ee{\end{equation}}
\def\ben{$$} \def\een{$$}
\def\ba{\begin{array}{c}}

\def\ea{\end{array}} 
\def\bea{\begin{eqnarray}}
\def\eea{\end{eqnarray}}

\def\ben{\begin{displaymath}}
\def\een{\end{displaymath}}
\def\ba{\begin{array}{c}}
\def\bal{\begin{array}{l}}
\def\ea{\end{array}}

\def\obj(#1)(#2)#3#4#5{%
  \psline[arrows={[-]}, linestyle=dashed, dash=0.0 1,dashadjust=false](#1)(#2)%
  \uput{0.4}[90](#1){#3}%
  \uput{0.4}[-90](#1){#4}\uput{0.4}[-90](#2){#5}%
}
\def\objd(#1)(#2)#3#4#5{%
  \psline[arrows={[-}, linestyle=dashed, dash=0.0 1,dashadjust=false](#1)(#2)%
  \uput{0.4}[90](#1){#3}%
  \uput{0.4}[-90](#1){#4}\uput{0.4}[-90](#2){#5}%
}
\def\theequation{\thesection.\arabic{equation}}

\begin{document}

\title{Hidden superconformal symmetry of spinless Aharonov-Bohm system}

\author{\textsf{Francisco Correa$^{a,b}$}\textsf{$\,$, Horacio Falomir$^c$}
\textsf{$\,$, V\'{\i}t Jakubsk\'y$^{a,d}$}\textsf{$\,$and  Mikhail S. Plyushchay$^a$}
\\
{\small \textit{${}^a$ Departamento de F\'{\i}sica, Universidad de Santiago
de Chile, Casilla 307, Santiago 2,
Chile  }}\\
{\small \textit{${}^b$ Centro de Estudios Cient\'{\i}ficos (CECS), Valdivia, Chile  }}\\
{\small \textit{${}^c$ IFLP/CONICET   -   Departamento   de   F\'{\i}sica,
 Facultad de Ciencias Exactas,}}\\
 {\small \textit{Universidad Nacional de La Plata, C.C. 67, (1900)  La Plata, Argentina}}\\
 {\small \textit{${}^d$ Nuclear Physics Institute, ASCR, 250 68 \v Re\v z, Czech Republic  }}\\
\sl{\small{E-mails:  fco.correa.s@gmail.com, falomir@fisica.unlp.edu.ar, v.jakubsky@gmail.com,
mplyushc@usach.cl}}}
\date{}
\maketitle

\begin{abstract}
A hidden supersymmetry is revealed in the spinless Aharonov-Bohm
problem. The intrinsic supersymmetric structure is shown to be
intimately related  with the scale symmetry.  As a result, a
bosonized superconformal symmetry is identified in the system.
Different self-adjoint extensions of the Aharonov-Bohm problem are
studied in the light of this superconformal structure and
interacting anyons. Scattering problem of the original
Aharonov-Bohm model is discussed in the context of the revealed
supersymmetry.
\end{abstract}

\section{Introduction}

The Aharonov-Bohm (AB) effect was discovered theoretically fifty
years ago \cite{aharonov,AB2}. Since that time it found various
experimental confirmations \cite{tonomura}, and has been
transformed into one of the most studied problems in planar
physics \cite{Schulman,WuYang,Schonf,Gold,Ruij,Berry,PH}; for a
nice review we refer the reader to \cite{reviewAB}. The AB effect
underlies the dynamical realization of anyons
\cite{LeinaasMyrheim,wilczek,anyonsbooks}, which currently are
supposed to play the key role in fractional Hall effect
\cite{Jain}. It appears in the analysis of cosmic strings
\cite{cosmicstrings,SGer}, and planar gravity \cite{GerJack,
planargravity}. This effect plays also important role in the
physics of graphene and nanotubes \cite{nanotubes,TD, JackiwAB}.

In their original work \cite{aharonov}, Aharonov and Bohm pointed
out the importance of the vector potential in quantum theory. Unlike
in classical mechanics, it has direct impact on the quantum dynamics
even when the electromagnetic field vanishes everywhere in the
regions accessible for a charged particle. Such a situation is
realized when the magnetic flux penetrating perpendicularly the
plane is contained in finite regions bounded by impenetrable
barrier. As a limit case we can consider the model given by the
vector potential
\begin{equation}\label{1-1}
    A^i= \frac{\alpha}{\hbar e}\epsilon^{ij}\frac{r^j}{\vec{r}{}\,{}^2}\,,
    \quad
    \vec{r}=(x,y),
\end{equation}
which corresponds to a singular flux that punctures the plane in the
origin $x=y=0$. In comparison with the free particle on the
punctured plane, the physics is changed via a nontrivial phase that wavefunction acquires when is moving around the point where the flux dwells. This is the core of the AB effect.

In this work, we are going to testify this model on the presence of
a \emph{hidden supersymmetry} \cite{BosoSusy}.
We will show that the Hamiltonian  of
a \textit{spinless} charged particle
 moving in presence of the vector potential
 (\ref{1-1})~\footnote{We choose units in which particle's mass $m=1/2$ and $\hbar=c=e=1$.},
\begin{equation}\label{1-3}
    {H}_{\alpha} = {{\mathcal{P}}_x}^{2} + {{\mathcal{P}}_y}^{2}=
    -\partial_r^2-\frac{1}{r}\partial_r+\frac{1}{r^2}\left(-i\partial_{\varphi}+\alpha\right)^2,
\end{equation}
\begin{equation}\label{1-2}
    {\mathcal{P}}_x = - i\, \partial_x -\alpha \, \frac{y}{r^2}\, ,
    \quad {\mathcal{P}}_y = - i\, \partial_y +\alpha \, \frac{x}{r^2},
\end{equation}
$x=r\cos\varphi$, $y=r\sin\varphi$, possesses a rich
algebraic structure of both exact (not dependent
on time explicitly) and dynamical (time dependent) integrals of
motion, that close for a superconformal superalgebra.

The key ingredients of a supersymmetric structure are supercharges
${Q}_a$, Hamiltonian ${H}$, and  a grading operator $\Gamma$. The
grading operator separates the set of relevant operators into
families of bosonic and fermionic observables in accordance with
whether they commute or anticommute with it. Supercharges are
supposed to be fermionic while Hamiltonian is the bosonic
operator,
\begin{equation}\label{Gamma}
    \{\Gamma,{Q}_a\}=[{H},\Gamma]=0\,,
    \quad \Gamma^2=1\,.
\end{equation}
We speak about hidden supersymmetry when the operators ${Q}_a$ and
$\Gamma$ can be found despite the  lack of fermionic (spin)
degrees of freedom in a system. The hidden supersymmetric
structure has been observed in various physically interesting
one-dimensional models,  including the Dirac delta function
potential problem, the reflectionless P\"oschl-Teller system
\cite{hiddensusy}, and  periodic finite-gap quantum systems
\cite{Lame,finitegap}. It was also observed in the bound state
Aharonov-Bohm effect \cite{hiddensusy}, that corresponds to a
particle confined to a circle.  In those systems, the hidden
supersymmetry reflects their peculiar spectral and scattering
properties.

We will seek for the operators $\Gamma$, ${Q}_1$ and
${Q}_2$ that would satisfy (\ref{Gamma}) and
\begin{equation}\label{1-8}
    \left\{{Q}_a , {Q}_b\right\} = 2 \delta_{a b}
    {H}_{\alpha}\,,
    \qquad \left[{H}_{\alpha} , {Q}_a \right] =0 \,,
    \quad {Q}_a={Q}_a^{\dagger},\ \ a,b=1,2\,.
\end{equation}
These relations correspond to Lie superalgebra of quantum
mechanical $N=2$ supersymmetry\footnote{In some systems, hidden
supersymmetry appears in a nonlinear form \cite{parabosonic}, in
which the anticommutator of supercharges is a polynomial in
Hamiltonian \cite{Andrianov}.}. The supercharges ${Q}_1$ and
${Q}_2$ can be nonlocal in general, as they correspond to the
square roots of the spinless differential operator $H_{\alpha}$.

The Hamiltonian ${H}_{\alpha}$ does not determine the dynamics of
the particle uniquely until its actual domain of definition is
fixed. The ambiguity in the proper definition of the system is
intimately related to the self-adjoint extensions of the
Hamiltonian. Physically, this corresponds to different
possibilities to realize  the condition of impenetrability of the
region $x=y=0$. The task of self-adjoint extensions has been
analyzed extensively in the literature. The case of a single
magnetic vortex has been studied as a limit case of an
impenetrable tube of finite radius with internal magnetic field
\cite{moroz}. It was also analyzed directly with making use of the
von Neumann theory of self-adjoint extensions
\cite{R-S,SGer,stovicek}.

Having in mind our objective, we cannot use these results directly
as they do not contain any information on the existence of the
supersymmetric structure described by (\ref{Gamma}), (\ref{1-8}).
Our approach will be different: we will identify first grading
operator $\Gamma$, and construct  operators ${Q}_1$ and ${Q}_{2}$
that will satisfy (\ref{Gamma}) and (\ref{1-8}) formally. Then we
will find their self-adjoint extensions. Hamiltonian, defined as the
square of supercharges, will be self-adjoint by construction
\cite{FalPis}. The obtained results will be compared with the known
ones. As we will see, the self-adjoint extension with regular wave
functions at the origin will be unitarily equivalent to the free
particle system for integer values of the magnetic flux, meanwhile
it will match exactly with the model discussed by Aharonov and Bohm
for non-integer values of $\alpha$. We also find two other
self-adjoint extensions of $H_\alpha$, which for non-integer values
of $\alpha$ possess hidden supersymmetry and correspond to
supersymmetric two-anyon systems with contact interaction.

The work is organized as follows. In the next Section, we
construct a formal supercharge that satisfies the required
properties. We then specify its self-adjoint extensions,
and discuss the existence of $N=2$ supersymmetry in the
system. Finding  the eigenfunctions of the associated
Hamiltonian, we show that the obtained system coincides
with the original model discussed by Aharonov and Bohm. We
analyze the action of supercharges on the wave functions to
clarify whether we have exact or spontaneously broken
supersymmetry. In Section \ref{exotic models}, we consider
other two self-adjoint extensions of $H_{\alpha}$, which
posses hidden supersymmetry. A particular attention is
given to the case of semi-integer flux in Section
\ref{semi-integer}, where an $su(2)$ family of grading
operators exists. In Section \ref{dynamical symmetries}, we
discuss conformal symmetry of the systems and confirm their
scale invariance. Sequently, we extend the algebraic
structure of the hidden supersymmetry by conformal
symmetry. In Section \ref{sectionanyons}, we provide an
alternative interpretation of the model in terms of anyons.
The last section is devoted to a brief summary and
discussion of the results, with emphasis on their physical
aspects. Particularly, we discuss the scattering problem in
the original Aharonov-Bohm model in the light of the hidden
supersymmetry and related translation symmetry breaking. We
list also there some open problems to be interesting for a
future research. Appendices include details on self-adjoint
extensions of the supercharges considered in Sections
\ref{ABSUSY} and \ref{exotic models}, and explicit formulas
for the domains of the operators discussed in Section
\ref{dynamical symmetries}.

\section{Hidden N=2 supersymmetry in spinless AB system\label{ABSUSY}}

In general, a formal Hamiltonian operator $H_{\alpha}$ (\ref{1-3})
admits a four-parametric  $U(2)$ family of self-adjoint extensions,
which specify physically different configurations, distinct in their
spectral and scattering properties \cite{stovicek,soldati}. The
spectrum depends strongly on the actual choice of the domain of
definition of $H_{\alpha}$ ; besides a continuous part of
non-negative energy scattering states, it may contain up to two
bound states of negative energy. As we stated above, our goal is to
examine the model for the presence of the hidden  supersymmetry
(\ref{Gamma}),  (\ref{1-8}) generated by self-adjoint supercharges.
This excludes immediately those self-adjoint extensions of
(\ref{1-3}) in which bound states are present, since negative energy
levels would imply purely imaginary eigenvalues for the
supercharges.

The general solution of the partial-wave stationary Schr\"odinger
equation for non-negative energy $E=k^2$, $k\geq0$,
\begin{equation}
    \label{timeH}
     H_{\alpha}\Psi_{k,l}=k^2\Psi_{k,l},
\end{equation}
is a linear combination of Bessel, ${\cal J}_{|l+\alpha|}(kr)$, and
Neumann, ${\cal Y}_{|l+\alpha|}(kr)$, functions multiplied by
$e^{il\varphi}$. The concrete choice of the linear combination is
specified uniquely by the domain of definition of the Hamiltonian.
In their seminal work
 \cite{aharonov}, Aharonov and Bohm considered the model where regular
at $r=0$  solutions were allowed only, i.e. their solution of
(\ref{timeH}) was of the form
\begin{equation}\label{ABreg}
    \Psi_{k,l}\sim
    {\cal J}_{|l+\alpha|}(kr)e^{il\varphi}.
\end{equation}
This gives rise to a unique fixing of the self-adjoint extension  of
the operator $H_{\alpha}$ that corresponds to the Aharonov-Bohm
system, which we denote by $H_{\alpha}^{AB}$.

The aim of the present section is to reveal a hidden supersymmetry
in the Aharonov-Bohm system. We proceed as follows: first of all, we
identify the $\Z_2$-grading operator of the bosonized supersymmetry.
Then we define a formal supercharge operator, find its self-adjoint
extension, and obtain the second odd generator of the $N=2$
supersymmetry. After that we show that the square of the found
supercharges coincides with the Hamiltonian $H_{\alpha}^{AB}$ of the
Aharonov-Bohm system.

\vskip0.2cm

Consider a nonlocal  operator of rotation in $\pi$,
\begin{equation}\label{1-6}
    {\cal R}\, f(x,y) = f(-x,-y)\, ,\qquad \mbox{or}\qquad {\cal R}\,
    f(r,\varphi) = f(r,\left.\varphi+\pi\right)\, ,
\end{equation}
which is presented in terms of the total angular momentum
$J=-i\partial_\varphi +\alpha$ as
\begin{equation}\label{RJ}
    {\cal R}=e^{-i\alpha\pi}e^{i\pi J}\, .
\end{equation}
It is a unitary, Hermitian involutive operator, ${\cal R}^2=1$,
which commutes with Hamiltonian (\ref{1-3}), and can be
identified as the grading operator $\Gamma$. Consider  a formal
nonlocal differential operator
\begin{equation}\label{QAB}
    Q_{\alpha}=\mathcal{P}_x+i{\cal R}(\alpha){\mathcal{P}_y}\, ,\quad
    \mbox{where}
     \quad {\cal R}(\alpha)=\left\{\begin{array}{ll}{\cal R},&
     \alpha\in (-1,0)\ \mbox{mod}\ 2\, ,\\
    {\cal R},& \alpha\in \mathbb{Z}\, ,\\
     -{\cal R},& \alpha\in (0,1)\ \mbox{mod}\ 2\, .
\end{array}\right.
\end{equation}
This operator and operator $i{\cal R}Q_{\alpha}$ satisfy formally
relations
\begin{equation}\label{QABR}
     \{Q_{\alpha},{\cal R}\}=\{i{\cal R}Q_{\alpha},{\cal R}\}=0\,,\quad
     \qquad\{Q_{\alpha},i{\cal R}Q_{\alpha}\}=0\,.
\end{equation}
On the other hand, we have
 \begin{equation}\label{QABH}
   \{Q_{\alpha},Q_{\alpha}\}=\{i{\cal R}Q_{\alpha},i{\cal R}Q_{\alpha}\}=
   2H_{\alpha}+2i{\cal R}[\mathcal{P}_x,\mathcal{P}_y]\,.
 \end{equation}
The commutator $[\mathcal{P}_x,\mathcal{P}_y]$ is just the
two-dimensional Dirac delta function. Unlike the one-dimensional
case, such a  term is not uniquely defined in the planar quantum
systems \cite{Albeverio, Bergman}. As it was discussed in
\cite{Jackiwdelta}, the self-adjoint extension of Hamiltonian
$H_{\alpha}$ has to be specified to define consistently the
operator. When we specify the actual domain of the self-adjoint
extension of $H_{\alpha}$, the Dirac delta function term is
redundant in the potential since its manifestation can be
understood in asymptotic behavior of the wave functions near the
origin.\footnote{ The same happens also in one dimension: when we
require the wave function to be continuous at $x=0$  and specify
its finite derivative jump there, the delta potential term can be
omitted from the Hamiltonian operator \cite{Albeverio}.} In our
current case, it will suffice to fix the self-adjoint extension of
$Q_{\alpha}$ since the square of self-adjoint operator is
self-adjoint as well.

Before we step to the analysis of the self-adjoint extension of
$Q_{\alpha}$, let us note that the actual choice of the signs  in
definition of ${\cal R}(\alpha)$ in (\ref{QAB}) is crucial.
Alternative choice of the sign for the same flux value case leads to
a different self-adjoint extension of $H_\alpha$, and will be
discussed in the next section. As we will see later in this section,
the exception is the case of integer flux values. For $\alpha\in
\Z$, both choices ${\cal R}(\alpha)={\cal R}$ and ${\cal
R}(\alpha)=-{\cal R}$ lead to the same result.

Operator $Q_{\alpha}$ (\ref{QAB}) defined on the smooth functions
with compact support is symmetric. Hence, the machinery of von
Neumann theory can be applied to find its self-adjoint extensions.
It can be checked that $Q_{\alpha}$ is essentially self-adjoint for
any $\alpha \in \R$. Indeed, the equations
$(Q_{\alpha})^{\dagger}f(r,\varphi)=\pm i f(r,\varphi)$ do not have
square integrable in $\R^2$ solutions. The deficiency index is equal
to $(0,0)$, and the operator $Q_{\alpha}$ does have a unique
self-adjoint extension, its closure, which we denote as
$Q_{\alpha}^{AB}$. Its domain of definition
$\mathcal{D}(Q_{\alpha}^{AB})$ is given by Eqs. (\ref{DQAB}) and
(\ref{DQABZ}) in  Appendix A.

To play the role of the supercharge, the operator $Q_{\alpha}^{AB}$
has to anticommute with the grading operator. The operator ${\cal
R}$ is essentially self-adjoint on $\mathcal{D}(Q_{\alpha}^{AB})$
and leaves this space invariant. Hence, the anticommutation relation
$\{Q_{\alpha}^{AB},{\cal R}\}=0$ is well defined on
$\mathcal{D}(Q_{\alpha}^{AB})$. This allows us to  construct
immediately the second self-adjoint supercharge $i{\cal
R}{Q}_{\alpha}^{AB}$, defined on $\mathcal{D}({Q}_{\alpha}^{AB})$ as
well. The square of the supercharges gives the self-adjoint
Hamiltonian $H_{\alpha}^{c}$ that is defined as
\begin{equation}\label{Hdom}
    H_{\alpha}^c=({Q}_{\alpha}^{AB})^2,
    \qquad\mathcal{D}(H_{\alpha}^c):=\left\{\Phi\in
    \mathcal{D}({Q}_{\alpha}^{AB})\, \vert\, {Q}_{\alpha}^{AB}\Phi \in
    \mathcal{D}({Q}_{\alpha}^{AB}) \right\}\,.
\end{equation}

Let us show now that the system described by $H_{\alpha}^{c}$
coincides with the model proposed by Aharonov and Bohm.  To do
this, we shall find eigenfunctions of $H_{\alpha}^c$.

To simplify the forthcoming analysis, let us comment  on relation
between the systems $H_{\alpha}^c$  and $H_{{\alpha+n}}^c$ with
magnetic flux values different in integer number $n\in\Z$.  A simple
formal operator equality
\begin{equation}\label{h}
    H_{\alpha+n}=U^{-1}_nH_{\alpha}U_{n}
\end{equation}
suggests that the unitary transformation  $U_n=e^{in\varphi}$ is
associated with the change of the magnetic flux of the system. It is
indeed the case.  First, we have $U^{-1}_{n}{\cal R}
U_{n}=(-1)^n{\cal R}$, and  for $\alpha\notin \Z$ there holds
\begin{equation}
    U^{-1}_{n}Q_{\alpha}^{AB}U_{n}=Q_{\alpha+n}^{AB}\,,\qquad
    U^{-1}_{n}\mathcal{D}(Q_{\alpha}^{AB})=\mathcal{D}
    (Q_{\alpha+n}^{AB})\,.
\end{equation}

The case of $\alpha\in \Z$ has, however,  a peculiarity, and deserves
a separate comment. For $\alpha=n$, there exists a system with
hidden supersymmetry represented by self-adjoint operators
$H^{c}_{n}$ and $Q_{n}^{AB}$ defined on corresponding domains. We
can use the transformation $U_1$ to construct another system with
the same flux, described by $H_{n}^{c}=U^{-1}_1H^{c}_{n-1}U_1$ and
$U_{1}^{-1}Q_{n-1}^{AB}U_1$. For the transformed supercharge domain,
there holds a relation
$$
    U_{1}^{-1}\mathcal{D}(Q_{n-1}^{AB})=\mathcal{D}
    (Q_{n}^{AB})\, ,
$$
which means that the independent integrals of motion $Q_{n}^{AB}$,
${\cal R}$ and $U_{1}^{-1}Q_{n-1}^{AB}U_1=\mathcal{P}_x-i {\cal
R}\mathcal{P}_y$ coexist in the same domain $\mathcal{D}
    (Q_{n}^{AB})$. Their linear
combinations (including their multiplications by ${\cal R}$) lead to
another set of integrals of motion, given by ${\cal R}$,
$\mathcal{P}_x$, $\mathcal{P}_y$ and their multiples by ${\cal R}$.
But $\mathcal{P}_x$ and $\mathcal{P}_y$  are the generators of
translation in the plane,  and, hence, the system described by
$H_{n}^{c}$ has a translational symmetry. As we will see, $H_0^{c}$
corresponds to a free particle in the plane, which, of course,
possesses translational invariance. Then, the revealed translational
symmetry of $H_{n}^{c}$ can be understood as a consequence of the
unitary equivalence of $H_0^{c}$ and $H_{n}^c$.

 Note that
 the Hamiltonian $H_{0}^{c}$ is invariant, in addition, under spatial
 reflections. As there is no preferential
 direction in the plane,
 we can consider two reflections
\begin{equation}\label{XYreflex}
  \mathcal{R}_xg(x,y) \mathcal{R}_x=g(-x,y)\,,\qquad \mathcal{R}_yg(x,y)
  \mathcal{R}_y=g(x,-y)\, ,
\end{equation}
which satisfy the  relations
\begin{equation} \label{RXY}
    \mathcal{R}_x^2=
     \mathcal{R}_y^2=1\,,\qquad
     [\mathcal{R}_x,\mathcal{R}_y]=0\,,\qquad
      \mathcal{R}=\mathcal{R}_x\mathcal{R}_y
     \, .
\end{equation}
In the polar coordinates their action is given by
\begin{equation}\label{Rpolar}
    \mathcal{R}_xf(r,\varphi)\mathcal{R}_x=f(r,\pi-\varphi)\,,\qquad
    \mathcal{R}_yf(r,\varphi)\mathcal{R}_y=f(r,-\varphi)\,.
\end{equation}
They commute with the operator ${\cal R}$, therefore they should be treated as nonlocal \emph{even} integrals of motion within the supersymmetric structure. Despite their involutive nature, neither of
these two operators can be identified as the grading operator since
they do not anticommute with the supercharge (\ref{QAB}) [they do
not commute with (\ref{QAB}) either]. As we shall see in Section
\ref{semi-integer}, the twisted analogs of the operators
(\ref{XYreflex}) emerge nontrivially in the systems with
half-integer flux.

We conclude that the change of the sign of ${\cal R}(\alpha)$ in
definition (\ref{QAB}) for integer flux value case reduces to a
unitary transformation, and that this sign ambiguity gives rise to
the translational invariance of $H_{n}^c$. At the same time, we
can see that the complete knowledge of the system for $\alpha\in
[-1,0)$ (or for $\alpha\in [0,1)$) provides a detailed description
for any other value of the magnetic flux as well. We will employ
this fact in the forthcoming analysis of the spectral properties
and supersymmetric structure of the system.

Let us fix the flux to be $\alpha\in[-1,0)$. In the polar
coordinates, the supercharge $Q_{\alpha}^{AB}$ reads
\begin{equation}\label{2-3}
    \begin{array}{c}
      {Q}_{\alpha}^{AB}
    =-i\, e^{i \varphi} \left[
    \partial_r - \frac{1}{r} \left(-i \partial_\varphi +\alpha \right)
    \right]\Pi_-\,
         -i\, e^{-i \varphi} \left[
    \partial_r + \frac{1}{r} \left(-i \partial_\varphi +\alpha \right) \right]
    \Pi_+\, ,
    \end{array}
\end{equation}
where
\begin{equation}\label{Pi+-}
    \Pi_\pm=\frac{1}{2}(1\pm {\cal R})
\end{equation}
are the projectors on the subspaces of even ($\Pi_+$) and odd
($\Pi_-$) partial waves. It preserves subspaces $\mathcal{H}_l$,
 \begin{equation}\label{invariant}
    \mathcal{H}_l:=\mathcal{L}\left\{e^{i(2l-1)\varphi},
    e^{i 2l\varphi} \right\}
    \otimes \mathbf{L}_2\left(\mathbb{R}^+; r\, dr \right)\subset
    \mathbf{L}_2\left( \mathbb{R}^2 \right)\,,\quad l \in\mathbb{Z}\,,
\end{equation}
where $\mathcal{L}\left\{e^{i(2l-1)\varphi},e^{i
2l\varphi} \right\}$ is a linear space spanned by the indicated
vectors. Then, the eigenvalue problem can be solved separately in
each $\mathcal{H}_l$.

The equation
\begin{equation}\label{eigenvaleq}
    {Q}_{\alpha}^{AB}\Phi_{l,\lambda}=\lambda \Phi_{l,\lambda} \quad
    \mbox{for }\quad \Phi_{l,\lambda}=\phi_{2l}(r)
    e^{i2l\varphi}+\phi_{2l-1}(r)e^{i(2l-1)\varphi}
\end{equation}
is  rewritten with help of (\ref{2-3}) in the form
\begin{equation}\label{3-1}
    \begin{array}{c}
    \displaystyle
      \phi_{2l}'(r) + \frac{{2l}+\alpha}{r}\, \phi_{2l}(r) =
      i \lambda\, \phi_{2l-1}(r)\,,
      \\ \\
      \displaystyle
      \phi_{2l-1}'(r) + \frac{1-(2l+\alpha)}{r}\, \phi_{2l-1}(r) =
      i \lambda\, \phi_{2l}(r) \,.
    \end{array}
\end{equation}
The general solution of (\ref{3-1}) for nonzero eigenvalues
$\lambda$ are linear combinations of the Bessel functions of the
first, $\mathcal{J}_{\nu}(|\lambda| r)$, and  second,
$\mathcal{Y}_{\nu}(|\lambda| r)$, kinds. The first is regular
while the other one is singular at the origin, but both are not
normalizable. To keep their interpretation in terms of
scattering states, we require the wavefunctions not to have too
strong divergence at
infinity\footnote{The mathematical framework for scattering states
is provided by the rigged Hilbert space, where the functions can
diverge at most as powers of $r$  \cite{rigged}.} and to
respect the behavior near the origin,
prescribed by the domain of definition. Since the
singular solution violates the first requirement due to its
divergence at $r=0$, it has to be discarded. Then the acceptable
solutions of (\ref{eigenvaleq}) for $\lambda\neq 0$ are
\begin{equation}\label{lnozeroscat}
 \Phi_{\lambda,l}\sim \mathcal{J}_{|2l+\alpha|}
 (|\lambda|r)e^{i2l\varphi}-i\frac{|\lambda|}{\lambda}
 \left\{\begin{array}{cll}
\mathcal{J}_{|1-2l-\alpha|}(|\lambda|r)e^{i(2l-1)\varphi}&
\mbox{for}& 2l+\alpha>0\, ,\\
-\mathcal{J}_{|1-2l-\alpha|}(|\lambda|r)e^{i(2l-1)\varphi}&
\mbox{for}& 2l+\alpha\leq0\, .
\end{array}\right.
\end{equation}

The solutions of the equations (\ref{3-1}) for $\lambda=0$ with
admissible behavior at the origin are
\begin{equation}\label{lnozerolambdazero}
 \Phi_{0,l}\sim\left\{\begin{array}{l}r^{2l+\alpha-1}e^{i(2l-1)\varphi}
 \quad \mbox{for}\quad 2l+\alpha\geq1\, ,\\
r^{-2l-\alpha}e^{i2l\varphi}\quad \mbox{for}\quad 2l+\alpha\leq0\, .
 \end{array}\right.
\end{equation}

We pass now to the analysis of the eigenfunctions of $H_{\alpha}^c$.
Hamiltonian commutes with the generator of rotations since
$\mathcal{D}(H_{\alpha}^c)$ is invariant with respect to the action
of $J$. Hence, one can find their common eigenfunctions
$\Psi_{|\lambda|,j}$,
\begin{equation}
     H_{\alpha}^c\Psi_{|\lambda|,j}=\lambda^2\Psi_{|\lambda|,j}\,,
     \quad J\Psi_{|\lambda|,l}=(l+\alpha)\Psi_{|\lambda|,l}\,.
\end{equation}
They can be composed from the eigenvectors of ${Q}_{\alpha}^{AB}$
corresponding to different signs of $\lambda$,
\begin{eqnarray}\label{scattering1}
     \Psi_{|\lambda|,2l}&\sim&\Phi_{\lambda,l}+\Phi_{-\lambda,l}
     \sim \mathcal{J}_{|2l+\alpha|}(|\lambda|r)e^{2il\varphi} ,\nonumber\\
    \Psi_{|\lambda|,2l-1}&\sim&\Phi_{\lambda,l}-\Phi_{-\lambda,l}
    \sim \mathcal{J}_{|1-2l-\alpha|}(|\lambda|r)e^{i(2l-1)\varphi}.
\end{eqnarray}
The zero-energy eigenstates of
$H_{\alpha}^{c}$ are
\begin{equation}\label{zeroE}
    \Psi_{0,l}\sim r^{|l+\alpha|}e^{il\varphi}.
\end{equation}

Note that the wave functions (\ref{scattering1}) vanish at the
origin except the special case of integer flux such that
$2l+\alpha=\beta\in\{0,1\}$. In this case,
$\mathcal{J}_{|\beta-2l-\alpha|}(|\lambda|r)=\mathcal{J}_{0}(|\lambda|r)\rightarrow
1$ for $r\rightarrow 0$, that is in agreement with the results on
the self-adjoint extension of the free particle in the punctured
plane \cite{kowalski}. The exclusion of the origin is of no
importance here since the considered functions are regular at this
point. In fact, the considered self-adjoint extension $H_{0}^{c}$ of
$H_{\alpha}$  with $\alpha=0$ is in correspondence with the system
of the free particle, since its domain of definition is spanned by
the same complete basis of partial waves
$\mathcal{J}_{|m|}(kr)e^{im\varphi}$.

We can compare the system represented by $H_{\alpha}^c$ with the
original setting of Aharonov and Bohm in the similar vein. The
behavior of the wave functions near the origin is prescribed in
the same way in both systems.  This leads to the same complete
basis of partial waves given by (\ref{lnozeroscat}),
(\ref{lnozerolambdazero}). Hence, $H_{\alpha}^c$ and
$H_{\alpha}^{AB}$ represent the same self-adjoint extension of
$H_{\alpha}$.

Thus,  the system described by $H_{\alpha}^c$ coincides with that
discussed originally by Aharonov and Bohm,
\begin{equation}
    H_{\alpha}^c=H_{\alpha}^{AB}\, .
\end{equation}
This means that the Aharonov-Bohm model possesses the hidden $N=2$
supersymmetry generated by the supercharges $Q_{\alpha}^{AB}$ and
$i{\cal R}Q_{\alpha}^{AB}$, in which the role of the grading
operator is played by the  operator ${\cal R}$. This result is
valid for any value of the magnetic flux.

Now, let us discuss the nature of the
revealed supersymmetry, and the action of the supercharges. The
spectrum of the operator $H_{\alpha}^{AB}$ consists of continuous
part only, which covers nonnegative real numbers. Any value of
energy $E$ is infinitely degenerate since there is an inifinite set
of linearly independent generalized wave functions
(\ref{scattering1}) corresponding to the given energy $E=\lambda^2$.
Let us discuss the action of the supercharges $Q_{\alpha}^{AB}$ and
$i{\cal R}Q_{\alpha}^{AB}$. The second supercharge interchanges the
eigenfunctions of $Q_{\alpha}^{AB}$ with different sign of
$\lambda\neq0$, i.e. there holds
 \begin{equation}\label{actQ}
     i{\cal R}Q_{\alpha}^{AB}
    \Phi_{\lambda,l}\sim\Phi_{-\lambda,l}\,.
\end{equation}
Consequently, with the direct use of this relation and
(\ref{scattering1}), we can write
 \begin{equation}\label{actH}
  Q_a
    \Psi_{|\lambda|,2l}\sim\Psi_{|\lambda|,2l-1},\quad  Q_a
    \Psi_{|\lambda|,2l-1}\sim\Psi_{|\lambda|,2l}\,,
 \end{equation}
where $Q_a$ is $Q_{\alpha}^{AB}$ or $i{\cal R}Q_{\alpha}^{AB}$.

\begin{figure}[h!]
\centering
\includegraphics[scale=0.835]{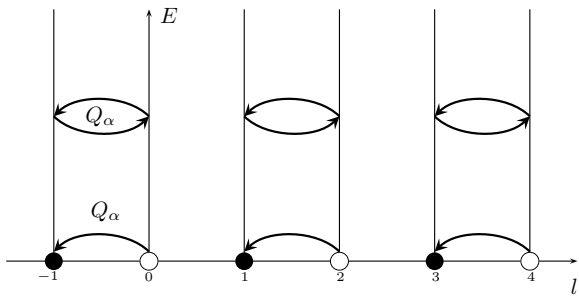}
\caption{For $\alpha\in[-1,0)\ \mbox{mod}\ 2$, the supercharges
$Q_{a}\in \{Q_{\alpha}^{AB},i\mathcal{R}Q_{\alpha}^{AB}\}$ preserve
the subspaces $\mathcal{H}_l$ defined in (\ref{invariant}).  We
illustrate the action of the supercharges in these subspaces for
$l=0,1,2$. The zero energy states (\ref{zeroE}) are represented by
the circles, the black circles correspond to the zero modes of
$Q_{a}$. The arrows between the same energy levels in each
$\mathcal{H}_l^{AB}$ correspond to relations (\ref{actH}) for  $E>0$
and to the relations (\ref{change2}) for $E=0$. } \label{susyex}
\end{figure}

The spectrum of $H_{\alpha}^{AB}$ includes infinitely degenerate
zero energy level.
We  restrict our consideration to the subspace $\mathcal{H}_l$ where
all the  energy levels are doubly degenerate. This subspace is
invariant under the action of the supercharges. Taking into account
Eq. (\ref{lnozerolambdazero}) for the zero modes of
$Q_{\alpha}^{AB}$, we conclude that there exists just a single state
in $\mathcal{H}_l$ annihilated by $Q_{\alpha}^{AB}$. Fixing $l\geq
0$, we can write explicit form of the involved functions,
\begin{equation}\label{change2}
    H_{\alpha}^{AB}\Psi_{0,2l}=H_{\alpha}^{AB}\Psi_{0,2l-1}=0\,,
    \quad Q_{\alpha}^{AB} \Psi_{0,2l}\sim\Psi_{0,2l-1}\,,\quad
    Q_{\alpha}^{AB}\Psi_{0,2l-1}=0\,.
\end{equation}
This resembles the Jordan blocks structure, which can appear in
diagonalization of a finite-dimensional matrix.
It does not contradict the self-adjointness of $Q_{\alpha}^{AB}$, the
supercharge can be diagonalized with making use of its eigenstates (\ref{lnozeroscat}) and
(\ref{lnozerolambdazero}). Hence, the supercharges annihilate just one half of the
zero-energy states. The rest of these states is transformed into
the kernel of the supercharges (see Fig.\ref{susyex}).

This picture can be compared with the cases of unbroken and broken
supersymmetry in non-periodic one-dimensional systems. There,
particularly, the unbroken supersymmetry is related to the
existence of a singlet bound state of zero energy, annihilated by
supercharge. The second, nonphysical solution corresponding to
zero energy is transformed to a physical one by a supercharge. In
the present case, the continuous nature of the spectrum  together
with the infinite degeneracy of the energy levels prevents us from
a similar classification of the revealed hidden supersymmetry. On
the other hand, there is some similarity of the revealed  hidden
supersymmetric structure with that appearing in one-dimensional
finite-gap periodic quantum systems, cf. \cite{finitegap}.

In conclusion of this section, let us make a few  comments on the
structure of the revealed supersymmetry, which further on will
provide an alternative interpretation of the system in terms of
anyons. In (\ref{Pi+-}) we introduced projectors $\Pi_{\pm}$ on
the subspaces of even and odd orbital angular momentum. This
allows us to  separate the domain $\mathcal{D}(H_{\alpha}^{AB})$
into two subsets $\Pi_{\pm}\mathcal{D}(H_{\alpha}^{AB})$, each of
which consists of eigenvectors of  $\mathcal{R}$ with fixed
eigenvalue $+1$ or $-1$. We can employ the matrix representation
of the projectors,
\begin{equation}\label{projector}
     \Pi_+=\left(\begin{array}{cc} 1&0\\0&0
             \end{array}
    \right),\quad \Pi_-=\left(\begin{array}{cc} 0&0\\0&1
             \end{array}
    \right).
\end{equation}
The Hamiltonian $H_{\alpha}^{AB}$ as well as other operators can be
rewritten in the matrix form,
\begin{equation}\label{matrixham}
      H_{\alpha}^{AB}=\left(\begin{array}{cc}H^{AB}_{\alpha,+}&0\\0&
         H^{AB}_{\alpha,-}\end{array}\right),\quad {\cal R}=
      \left(\begin{array}{cc}1&0\\0&-1\end{array}\right),
\end{equation}
where $H^{AB}_{\alpha,\pm}=\Pi_{\pm}H_{\alpha}^{AB}$. The
supercharge $Q_{\alpha}^{AB}$ is antidiagonal operator and its
explicit form for $\alpha\in[-1,0)$ can be deduced from
(\ref{2-3}) and (\ref{projector}). In this framework, the wave
function $\psi$ from the domain of $H_{\alpha}^{AB}$ is just a
column vector, whose upper element is composed of even partial
waves, $\psi_+=\Pi_+\psi=\sum_{l\in\Z}g^+_l(r)e^{i2l\varphi}$,
while the lower component consists of odd partial waves,
$\psi_-=\Pi_-\psi=\sum_{l\in\Z}g^-_l(r)e^{i(2l-1)\varphi}$. Such a
representation reveals an obvious similarity of the hidden
supersymmetry of the spinless Aharonov-Bohm system  with
supersymmetry of a usual form, associated with introduction of the
spin degrees of freedom \cite{Wit,CKS}.

\section{Exotic models\label{exotic models}}

The choice of the  signs we made in definition of ${\cal
R}(\alpha)$ in (\ref{QAB}), and the observed ambiguity for
$\alpha\in \Z$ case, led us to the revealing of the hidden $N=2$
supersymmetry in the original Aharonov-Bohm system. In this
section we investigate the consequences of the alternative choice
of the signs in (\ref{QAB}) for non-integer flux values.

So, let us consider the operator
\begin{equation}\label{Qgamma}
 \tilde{Q}_{\alpha}=\mathcal{P}_x-i{\cal R}(\alpha){\mathcal{P}_y},
 \quad \mbox{where}\quad {\cal R}(\alpha)=\left\{\begin{array}{ll}{\cal R}&
 \alpha\in (1,2)\ \mbox{mod}\ 2\, ,\\
 -{\cal R}& \alpha\in (0,1)\ \mbox{mod}\ 2\, .
\end{array}\right.
\end{equation}
The formal relations (\ref{QABR}) and (\ref{QABH}) imposed on the
supercharge remain intact, up to the sign of the commutator term
$[\mathcal{P}_x,\mathcal{P}_y]$ in the square of
$\tilde{Q}_{\alpha}$. This suggests that the difference, if any,
could appear in self-adjoint extensions of the supercharge operator
(\ref{Qgamma}).

The transformation $U_1$ changes the flux of the system in one
unit. It maintains the self-adjointness of the operators, i.e.
when an operator ${\cal O}$ is self-adjoint on $\mathcal{D}({\cal
O})$, the operator $\tilde{{\cal O}}=U_{1}^{-1}{\cal O}U_1$ is
self-adjoint on $U_1^{-1}\mathcal{D}({\cal O})$. This means that
when we find all the admissible self-adjoint extensions of
$\tilde{Q}_{\alpha}$ for $\alpha\in(-1,0)|_{mod\ 2}$, we can get
all the self-adjoint extensions of the operator for $\alpha\in
(0,1)|_{mod\ 2}$ just by application of this transformation. The
inverse  is also true with changing of the transformation $U_1$
for $U_1^{-1}=U_{-1}$. Without loss of generality, we restrict our
analysis to $\alpha\in(0,1)|_{mod\ 2}$.

As the operator $\tilde{Q}_{\alpha}$ for $\alpha\in(0,1)|_{mod\
2}$ coincides formally with the operator ${Q}_{\alpha}$ for
$\alpha\in(-1,0)|_{mod\ 2}$, we can use directly Eq. (\ref{2-3})
to express the operator in polar coordinates, just keeping in mind
the different range of $\alpha$. The operator $\tilde{Q}_{\alpha}$
preserves the subspaces (\ref{invariant}), and  is symmetric on
$C_0^{\infty}(\mathbb{R}^2-\{0\})$. The domains of its conjugate
and its closure are presented in Appendix A.

We have to solve the deficiency equations
$\tilde{Q}_{\alpha}^{\dagger}\psi=\pm i\psi$ to reveal the bases of
the deficiency subspaces. The relation (\ref{2-3}) together with
(\ref{invariant}) simplify this task since the problem can be
inspected for each subspace $\mathcal{H}_l$ separately. The
deficiency indexes are vanishing again in all the subspaces
$\mathcal{H}_{l}$ except the subspace $\mathcal{H}_{l_0}$ given by
the integer $l_0$ such that $2l_0+\alpha\in (0,1)$. In contrary to
(\ref{QAB}), the deficiency indexes of $\tilde{Q}_{\alpha}$ are
$(1,1)$, so that there exists a $U(1)$ family of self-adjoint
extensions $\tilde{Q}_{\alpha}^{\gamma}$ of $\tilde{Q}_{\alpha}$.
The detailed derivation of the result is rather technical (see
Appendix A), and we present the final form of the domain of the
self-adjoint operator $\tilde{Q}_{\alpha}^{\gamma}$\,:
\begin{equation}\label{4-1a}
    \begin{array}{c}
      \displaystyle
      \mathcal{D}\left( \tilde{Q}_{\alpha}^\gamma \right):=
      \left\{ \Phi(r,\varphi)=f(r,\varphi)+ A \left[\Phi_+(r,\varphi)
    + e^{i \gamma} \Phi_-(r,\varphi) \right]\, \vert\, \right.
      \\
      \displaystyle
      \left. f(r,\varphi)\in \mathcal{D}\left( \overline{Q_{\alpha}} \right)
      \,, A\in \mathbb{C} \,, \gamma\in [0,2\pi) \right\}\, ,
    \end{array}
\end{equation}
where $\overline{Q}_{\alpha}$ is the closure  of
$\tilde{Q}_{\alpha}$. Expanding the function $f(r,\varphi)\in
\mathcal{D}( \overline{Q_{\alpha}})$ in partial waves
$f(r,\varphi)=\sum_{l}f_l(r)e^{il\varphi}$, we find that the
radial parts $f_l(r)$ have to have  the following asymptotic
behavior near the origin: $|f_l(r)|= O(1)$ for
$l\notin\{2l_0,2l_0-1\}$, while $|f_{2l_0}(r)|=
o(r^{-2l_0-\alpha})$ and $|f_{2l_0-1}(r)|= o(r^{-1+2l_0+\alpha})$.
The functions $\Phi_{\pm}(r,\varphi)$ form the basis of deficiency
subspaces, $ \tilde{Q}_{\alpha}^{\dagger}\Phi_{\pm}(r,\varphi)=\pm
i \Phi_{\pm}(r,\varphi)$, and can be written in terms of McDonald
functions
\begin{equation}\label{defsubs}
    \Phi_\pm = 
    K_{2l_0+\alpha}( r)e^{2l_0i\varphi}
    \pm  K_{1-(2l_0+\alpha)}(r)e^{(2l_0-1)i\varphi}\,.
\end{equation}

The operator ${\cal R}$ is essentially self-adjoint on
$\mathcal{D}\left( \tilde{Q}_{\alpha}^\gamma \right)$, but the
requirement $\{\tilde{Q}_{\alpha}^{\gamma},{\cal R}\}=0$ is
consistent if and only if the operator ${\cal R}$ leaves
$\mathcal{D}(\tilde{Q}_{\alpha}^{\gamma})$ invariant. Using
(\ref{4-1a}) and the fact that ${\cal
R}\mathcal{D}(\overline{Q}_{\alpha})=\mathcal{D}(\overline{Q}_{\alpha})$,
we get
\begin{equation}\label{5-2}
    {\cal R}\big( \mathcal{D}(\tilde{Q}_{\alpha}^\gamma) \big) =
    \mathcal{D}(\tilde{Q}_{\alpha}^{2\pi-\gamma})\,.
\end{equation}
The requirement on the invariance of
$\mathcal{D}(\tilde{Q}_{\alpha}^\gamma)$ holds true for two values
of parameter $\gamma$ only,
\begin{equation}\label{gamma}
 \gamma=0,\pi \mod(2\pi).
\end{equation}
Hence, the $N=2$ supersymmetric structure is admissible just for
these values of the parameter $\gamma$. If not stated otherwise, we
will restrict $\gamma\in\{0,\pi\}$ from now on. In this case, the
domains of $\tilde{Q}^{\gamma}_{\alpha}$ and $i{\cal
R}\tilde{Q}^{\gamma}_{\alpha}$ coincide. It is worth to mention that
both $\mathcal{D}(\tilde{Q}_{\alpha}^{0})$ and
$\mathcal{D}(\tilde{Q}_{\alpha}^{\pi})$ acquire particularly simple
form,
\begin{equation}\label{dom0}
\begin{array}{c}
      \displaystyle
      \mathcal{D}\left( \tilde{Q}_{\alpha}^{0} \right):=
      \left\{ \Phi(r,\varphi)=f(r,\varphi)+ A\ K_{2l_0+\alpha}(r)
      e^{2il_0\varphi}\, |\, f(r,\varphi)\in \mathcal{D}\left(
\overline{Q_{\alpha}} \right),
      \,\, A\in \mathbb{C} \right\}\, ,
    \end{array}
\end{equation}
\begin{equation}\label{dompi}
\begin{array}{c}
      \displaystyle
      \mathcal{D}\left( \tilde{Q}_{\alpha}^{\pi} \right):=
      \left\{ \Phi(r,\varphi)=f(r,\varphi)+ A\ K_{1-2l_0-\alpha}(r)
      e^{(2l_0-1)i\varphi}\,|\, f(r,\varphi)\in \mathcal{D}\left(
\overline{Q_{\alpha}} \right)
      , \, \,A\in \mathbb{C} \right\}\, ,
    \end{array}
\end{equation}
which manifests their invariance with respect to rotations generated
by $J$.

The structure of $N=2$ supersymmetry is completed by the following
definition of the self-adjoint Hamiltonian $H^{\gamma}_{\alpha}$\,:
 \begin{equation}\label{4-9}
    H^\gamma_{\alpha}= \left({\tilde{Q}_{\alpha}^\gamma}\right)^2,\quad
    \mathcal{D}(H^\gamma_{\alpha}):=\left\{\Phi\in \mathcal{D}
    (\tilde{Q}_{\alpha}^\gamma) \,\vert\, \tilde{Q}_{\alpha}^\gamma
    \Phi \in \mathcal{D}(\tilde{Q}_{\alpha}^\gamma) \right\}\,.
\end{equation}
Hence, taking the different definition (\ref{Qgamma}) of the
supercharge, we reveal two distinct self-adjoint extensions
$H_{\alpha}^{\gamma}$ of the formal Hamiltonian operator
$H_{\alpha}$,  which, like the Aharonov-Bohm system considered in
the previous section, are characterized by  the hidden $N=2$
supersymmetry. As the domains of Hamiltonians are invariant with
respect to $J$, the systems have rotational symmetry as well.

In the next step we shall analyze the spectrum of
$H_{\alpha}^{\gamma}$ and find the associated wavefunctions. Since
$H^{\gamma}_{\alpha}$ is the square of the self-adjoint operator
$\tilde{Q}_{\alpha}^{\gamma}$, and
$\mathcal{D}(H^{\gamma}_{\alpha})$ is a subset of
$\mathcal{D}(\tilde{Q}_{\alpha}^{\gamma})$, we conclude that in
correspondence with the hidden supersymmetric structure, the
spectrum is non-negative.

We can employ Eqs. (\ref{2-3}), (\ref{eigenvaleq}) and (\ref{3-1}),
keeping in mind the different range of $\alpha$,
$2l_0+\alpha\in(0,1)$, $l_0\in \mathbb{Z}$. Singular solutions of
(\ref{3-1}) have to be discarded in the subspaces $\mathcal{H}_l$
for $l\neq l_0$. Hence, the eigefunctions
$\Phi_{\lambda,l}=\phi_{2l} e^{2il\varphi}+\phi_{2l-1}
e^{i(2l-1)\varphi}$ lying in these subspaces have exactly the same
form as (\ref{lnozeroscat}) and (\ref{lnozerolambdazero}). The
situation is different in the subspace $\mathcal{H}_{l_0}$. Due to
(\ref{4-1a}), the admissible solutions $\Phi_{\lambda,l_0}$ in
$\mathcal{H}_{l_0}$ have to fit the following asymptotic behavior
near the origin\,:
\begin{equation}\label{4-4}
    \begin{array}{c}
      \displaystyle
      \phi_{2l_0}(r)= A \left(1+e^{i \gamma} \right)
      \frac{\Gamma(2l_0+\alpha)}{2^{1- (2l_0+\alpha)}}\,
       r^{-2l_0-\alpha} +o\left( r^{-2l_0-\alpha} \right) \,,
      \\  \\
      \displaystyle
      \phi_{2l_0-1}(r)= A \left(1-e^{i \gamma} \right)
      \frac{\Gamma(1-(2l_0+\alpha))}{2^{2l_0+\alpha}}\,
       r^{-1+2l_0+\alpha} +o\left( r^{-1+2l_0+\alpha} \right)
       \,,
    \end{array}
\end{equation}
dictated explicitly by the relevant part
$A(\Phi_+(r,\varphi)+e^{i\gamma}\Phi_-(r,\varphi))$ of the domain of
$\tilde{Q}_{\alpha}^{\gamma}$, where $A$ is a constant. The
solutions of (\ref{3-1}) for $\lambda\neq 0$ are
 \begin{equation}\label{sol-1}
        \begin{array}{c}
      \displaystyle
      \phi_{2l_0}(r) = C_1 \mathcal{J}_{\left|2l_0+\alpha \right|}
      (|\lambda| r)+C_2 \mathcal{Y}_{\left|2l_0+\alpha \right|}(|\lambda| r)\,,
      \\ \\
      \displaystyle
      \phi_{2l_0-1}(r) =-i \frac{|\lambda |}{\lambda}
                \left(C_1 \mathcal{J}_{2l_0+\alpha-1}(|\lambda| r)
                +C_2 \mathcal{Y}_{2l_0+\alpha-1}(|\lambda| r)\right)\,,
        \end{array}
\end{equation}
with the coefficients related to $A$,
\begin{eqnarray}\label{4-7}
      \frac{C_2}{A} &=& -\frac{\pi}{2}\left( \frac{|\lambda|}
      {\mu} \right)^{2l_0+\alpha}
      \left( 1+e^{i \gamma} \right)\,,\nonumber\\
      \frac{C_1}{A} &=&
     \frac{\pi}{2\sin \left(\pi(2l_0+\alpha)\right)}
\left\{
     \frac{i\lambda}{|\lambda|}
     \left( \frac{|\lambda|}{\mu}
    \right)^{1-2l_0-\alpha}
     \left( 1-e^{i \gamma}
    \right) \right.\nonumber\\
     &&\left.+\cos \left(\pi(2l_0+\alpha)\right)
     \left( \frac{|\lambda|}{\mu} \right)^{2l_0+\alpha}
     \left( 1+e^{i \gamma} \right)
     \right\}.
\end{eqnarray}
The solution of (\ref{3-1}) for $\lambda=0$ reads
\begin{equation}\label{zerosolzero}
 \Phi_{l_0,0}\sim\left\{\begin{array}{lcl} r^{-2l_0-\alpha}
 e^{2il_0\varphi}&\mbox{for}& \gamma=0\, ,\\
                r^{-1+2l_0+\alpha}e^{(2l_0-1)i\varphi}
                &\mbox{for}& \gamma=\pi\, .\end{array}\right.
\end{equation}
\noindent

Likewise in the previous section, there holds
$[H_{\alpha}^{\gamma},J]=0$ since ${\cal D}(H_{\alpha}^{\gamma})$ is
invariant with respect to the action of $J$.  Hence, one can find
the common eigenfunctions $\Psi_{|\lambda|,j}$,
\begin{equation}
     H_{\alpha}^{\gamma}\Psi_{|\lambda|,j}=\lambda^2
     \Psi_{|\lambda|,j}\,,\qquad J\Psi_{|\lambda|,j}=(j+\alpha)\Psi_{|\lambda|,j}\,.
\end{equation}
They can be composed from the eigenvectors of
$\tilde{Q}_{\alpha}^{\gamma}$ which correspond to eigenvalues
$\pm\lambda$. As long as $l\neq l_0$, the scattering states of
$H_{\alpha}^{\gamma}$ take the form (\ref{scattering1}). For
$l=l_0$, we get
\begin{eqnarray}
    \Psi_{|\lambda|,2l_0}&\sim&\Phi_{\lambda,2l_0}+\Phi_{-\lambda,2l_0}
    \sim [\cos\pi\tilde{\alpha}
    \mathcal{J}_{\tilde{\alpha}}(|\lambda|r)
    -\sin\pi\tilde{\alpha}\mathcal{Y}_{\tilde{\alpha}}(|\lambda|r)]e^{2l_0i\varphi}\,
     ,\quad \gamma=0\,,\label{f1}\\
    \Psi_{|\lambda|,2l_0-1}&\sim&\Phi_{\lambda,2l_0}-\Phi_{-\lambda,2l_0}
    \sim \mathcal{J}_{1-\tilde{\alpha}}(|\lambda|r)e^{i(2l_0-1)\varphi},
    \quad \gamma=0\,,\label{f2}\\
    \Psi_{|\lambda|,2l_0}&\sim&\Phi_{\lambda,2l_0}-\Phi_{-\lambda,2l_0}
    \sim \mathcal{J}_{\tilde{\alpha}}(|\lambda|r)e^{2l_0i\varphi}\, ,
    \quad \gamma=\pi\,,\label{f3}\\
    \Psi_{|\lambda|,2l_0-1}&\sim&\Phi_{\lambda,2l_0}+\Phi_{-\lambda,2l_0}
    \sim\left[\cos\pi\tilde{\alpha}\, \mathcal{J}_{1-\tilde{\alpha}}
    (|\lambda|r)-\sin\pi\tilde{\alpha}\,\mathcal{Y}_{1-\tilde{\alpha}}
    (|\lambda|r)\right] e^{(2l_0-1)i\varphi},\quad \gamma=\pi\,,
\label{f4}
\end{eqnarray}
where $\tilde{\alpha}=2l_0+\alpha$.
Like in (\ref{actQ}), the second
supercharge $i{\cal R}\tilde{Q}_{\alpha}^{\gamma}$ interchanges
eigenvectors of $\tilde{Q}_{\alpha}^{\gamma}$ with differents signs
of $\lambda$. Consequently, the supercharges interchange the wave
functions $\Psi_{|\lambda|,2l}$ and $\Psi_{|\lambda|,2l-1}$ given by
(\ref{scattering1}) for $l\neq l_0$, and by (\ref{f1})-(\ref{f4})
for $l=l_0$. In contrary to the Aharonov-Bohm Hamiltonian
$H_{\alpha}^{AB}$, the operator $H_{\alpha}^{0}$ (resp.
$H_{\alpha}^{\pi}$) has a singular zero-mode
$\Psi_{0,2l_0}=r^{-2l_0-\alpha}e^{2l_0i\varphi}$ (resp.
$\Psi_{0,2l_0-1}=r^{-1+2l_0+\alpha}e^{(2l_0-1)i\varphi}$) in the
subspace $\mathcal{D}(H_{\alpha}^{0})\cap\mathcal{H}_{l_0}$ (resp.
$\mathcal{D}(H_{\alpha}^{\pi})\cap\mathcal{H}_{l_0}$)~\footnote{ The
zero modes of $H_{\alpha}^{\gamma}$ as well as $H_{\alpha}^{AB}$ can
be understood as a low energy limit of the properly normalized
scattering states.}. However, there are no other differences in the
analysis, the supercharge $Q_{\alpha}^{\gamma}$ annihilates just a
half of the zero-energy states, mapping the rest to its kernel.
Hence the action of the supercharge $Q_{\alpha}^{\gamma}$ is
qualitatively in the complete agreement with the discussion
presented for $Q_{\alpha}^{AB}$ in the previous section.

\begin{figure}[h!]
\centering
\includegraphics[scale=1.2]{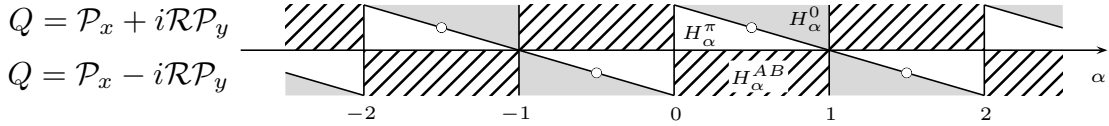}
\caption{The figure illustrates the three different self-adjoint
extentions of $H_{\alpha}$ in dependence on $\alpha$. Upper and
lower cases correspond to different definitions (\ref{QAB}) and
(\ref{Qgamma}) of the supercharges. Rectangular shaded zones
correspond to the setting discussed by Aharonov and Bohm, while
gray and white triangular zones correspond to the exotic models
represented by $H_{\alpha}^0$ and $H_{\alpha}^{\pi}$ respectively.
The circles for half-integer values of $\alpha$ indicate that
$H_{\alpha}^0$ and $H_{\alpha}^{\pi}$ are unitarily equivalent in
this case, see the next Section \ref{semi-integer}.}
\label{susyex1}
\end{figure}

\section{Half-integer flux and twisted reflections\label{semi-integer}}

The operator $\mathcal{R}$ commutes formally with $H_{\alpha}$ for
any value of the magnetic flux $\alpha$. In contrary, the reflection
operators ${\cal R}_x$ and ${\cal R}_y$ defined in (\ref{XYreflex})
are exclusive integrals of motion of the free particle. When the
magnetic flux is switched on, they provoke a change of the sign of the
magnetic flux in $H_\alpha$,  i.e. ${\cal R}_x H_\alpha {\cal
R}_x={\cal R}_yH_\alpha {\cal R}_y =H_{-\alpha}$.

We can define the ``twisted'' reflection operators $\tilde{{\cal
R}}_x=e^{i\alpha\pi}e^{-2i\alpha\varphi}{\cal R}_x$ and
$\tilde{{\cal R}}_y=e^{-2i\alpha\varphi}{\cal R}_x,$ for which
formally  $[H_{\alpha},\tilde{\cal R}_x ]=[H_{\alpha},\tilde{\cal
R}_y ]=0$ and
\begin{equation}\label{R2R2}
    \tilde{\cal R}_x^2=\tilde{\cal R}_y^2=1\, .
\end{equation}
For a general value of $\alpha$, however, they do not preserve the
space of $2\pi$-periodic functions. $\tilde{{\cal R}}_x$ and
$\tilde{{\cal R}}_y$ are defined consistently for $\alpha=m$ or
$\alpha=m+\frac{1}{2}$ only, where $m \in \mathbb{Z}$. For
$\alpha=m$, these operators are related to non-twisted reflections
(\ref{XYreflex}) by the unitary transformation $U_m=e^{im\varphi}$,
and they commute, therefore, with $\mathcal{R}$.  The situation is
essentially different for half-integer values of $\alpha$. For
$\alpha=m+\frac{1}{2}$ we get
\begin{equation}\label{twistXY}
     \tilde{\cal R}_x=-ie^{-i(2m+1)\varphi}{\cal R}_x,
     \quad \tilde{\cal R}_y=e^{-i(2m+1)\varphi}{\cal R}_y,
\end{equation}
and
\begin{equation}\label{Liedeform}
     [\tilde{\cal R}_x,\tilde{\cal R}_y]=2i {\cal R},
     \qquad [{\cal R},\tilde{\cal R}_x]=
     2i \tilde{\cal R}_y,\qquad [{\cal R},\tilde{\cal R}_y]
     =-2i \tilde{\cal R}_x\, ,
\end{equation}
where for the sake of convenience we included in definition of
$\tilde{\cal R}_x$ an additional  numerical factor $(-1)^{m+1}$. The
operators satisfy also
 \begin{equation}\label{alphapul}
  \{\tilde{\cal R}_x,\tilde{\cal R}_y\}=
  \{\tilde{\cal R}_x,{\cal R}\}=\{\tilde{\cal R}_y,{\cal R}\}=0
  \,.
 \end{equation}

Relations (\ref{Liedeform}), (\ref{alphapul}) mean that the
twisted reflection operators (\ref{twistXY}) together with ${\cal
R}$ satisfy exactly the same set of algebraic relations as the
three Pauli matrices, i.e. up to the numerical factor
$\frac{1}{2}$ they are generators of the spinorial representation
of $su(2)$. In this section, we discuss the role of the triplet of
reflection operators for the hidden supersymmetry of the systems
with half-integer flux~\footnote{The special ``magic" of half
fluxons was discussed in the context different from the present
one in \cite{BY,AhCol}.}.

\vskip0.1cm
 Without loss of generality, set $\alpha=1/2$.
The operators ${\cal R}$, $\tilde{\cal R}_x=-ie^{-i\varphi}{\cal
R}_x$ and $\tilde{\cal R}_y=e^{-i\varphi}{\cal R}_y$ are symmetric
on $\mathcal{D}(Q_{1/2}^{AB})$, or
$\mathcal{D}(\tilde{Q}_{1/2}^{\gamma})$, $\gamma\in[0,2\pi)$. In
addition, neither $\tilde{\cal R}_xf=\pm i f$ nor $\tilde{\cal
R}_yf=\pm i f$ have nontrivial solutions. The operators are
essentially self-adjoint both on $\mathcal{D}({Q}_{1/2}^{AB})$ and
$\mathcal{D}(\tilde{Q}_{1/2}^{\gamma})$. As $\tilde{\cal
R}^2_x=\tilde{\cal R}^2_y=1$, they are unitary as well. The
described properties of the triplet of reflection operators allows
us to introduce a three-parametric family of $SU(2)$-transformations
 \begin{equation}
    {\cal U}(\beta_x,\beta_y,\beta)=
     e^{i (\beta_x\tilde{\cal R}_x+\beta_y
     \tilde{\cal R}_y+\beta {\cal R})}\,, \label{U}
 \end{equation}
which will be important in the forthcoming analysis.

Consider now the Aharonov-Bohm model described  by $H_{1/2}^{AB}$.
The domain (\ref{DQAB}) of the supercharge $Q_{1/2}^{AB}$ is
invariant under the action of all the triplet of reflections,
\begin{equation}
     {\cal R}\mathcal{D}(Q_{\alpha}^{AB})=\tilde{\cal R}_x\mathcal{D}
     (Q_{1/2}^{AB})=\tilde{\cal R}_y\mathcal{D}(Q_{1/2}^{AB})
     =\mathcal{D}(Q_{1/2}^{AB})\,.
\end{equation}
Therefore, the set of integrals of motion of $H_{1/2}^{AB}$
consisting of $ Q_{1/2}^{AB},\ i{\cal R}Q_{1/2}^{AB}$, ${\cal R}$
and $J$ has to be extended by the operators $\tilde{\cal R}_{x}$ and
$\tilde{\cal R}_{y}$.

The question is then how they could be incorporated into the
superalgebraic structure of the system. Keeping ${\cal R}$ as the
grading operator, the new integrals of motion are of the fermionic
nature. Hence, the anticommutators with the supercharges
$Q_{1/2}^{AB}$ and $i{\cal R}Q_{1/2}^{AB}$ should be computed, as
well as their commutators with $J$. Both twisted reflections,
however, anticommute with angular momentum generator $J$. As a
consequence, the repeated commutators  with  $J$ give
\begin{equation}
     [\tilde{\cal R}_x,J]=2\tilde{\cal R}_xJ,\quad
      [\tilde{\cal R}_xJ,J]=2\tilde{\cal R}_xJ^2,
     \ldots,\quad [\tilde{\cal R}_xJ^n,J]=2\tilde{\cal
     R}_xJ^{n+1}\,,
\end{equation}
and analogous relations for $\tilde{\cal R}_y$. Subsequent
anticommutators of the odd integrals $\tilde{\cal R}_xJ^n$ and
$\tilde{\cal R}_yJ^k$, $n,k=0,1\ldots$, produce the  integrals of
the form ${\cal R}J^n$. In the same way, the  anticommutators of the
twisted reflections with the supercharges $Q_{1/2}^{AB}$ and $i{\cal
R}Q_{1/2}^{AB}$, and corresponding repeated (anti)commutation
relations reproduce the basic integrals multiplied by
$(H_{1/2}^{AB})^n$. We see that the inclusion of the twisted
reflection operators into the superalgebraic structure leads to its
nonlinear deformation characterized by appearance of the
multiplicative factors $(J)^n$ and $(H_{1/2}^{AB})^k$ in
(anti)commutation relations, cf. \cite{ hiddensusy, finitegap,
parabosonic,Andrianov}.

Since the integrals $\tilde{\cal R}_x$ and $\tilde{\cal R}_y$
satisfy the relations (\ref{R2R2}), any of them can also be taken
as the $\Z_2$-grading operator instead of ${\cal R}$. Such a
possibility for alternative choice of the grading operator
resembles the tri-supersymmetric structure studied in
\cite{finitegap}. The difference is that here  the three
involutive integrals mutually anticommute, while in the
tri-supersymmetric structure analogous integrals mutually
commute\footnote{A supersymmetric structure with three mutually
anticommuting involutive integrals of motion was observed recently
in Bogolyubov-de Gennes system \cite{BdG}.}. If, for instance,
$\tilde{\cal R}_x$ is identified as the grading operator, the
operators $Q_{1/2}^{AB}$ and $i\tilde{\cal R}_xQ_{1/2}^{AB}$ will
be nontrivial supercharges, and the angular momentum $J$ has also
to be treated as an odd generator. The anticommutator of the
supercharge $i\tilde{\cal R}_xQ_{1/2}^{AB}$ with ${\cal R}$
generates then $i\tilde{\cal R}_yQ_{1/2}^{AB}$, that has to be
treated  as an even integral. Further computing shows that with
${\cal R}_x$ taken as the grading operator, we have, again, a
nonlinear superalgebraic structure.

The picture  is completely different in the case of the systems
described by $H_{1/2}^{\gamma}$ ($\gamma\in \{0,\pi\}$). The
operators $\tilde{\cal R}_x$ and $\tilde{\cal R}_y$ are no longer
symmetries of the system as the domain
$\mathcal{D}(H_{\alpha}^{\gamma})$ is not invariant under their
action. The unitary transformations (\ref{U}) can be used to map the
system  $H_{1/2}^{\gamma}$ ($\gamma\in \{0,\pi\}$) to another,
equivalent one, with the same supersymmetric structure. Let us
discuss a few particular examples, where the transformed grading
operator acquires particularly simple form. We are interested in the
mappings which would interchange the operators ${\cal R}$,
$\tilde{\cal R}_x$ and $\tilde{\cal R}_y$ in the role of the grading
operator,
\begin{equation}\label{RU}
     {\cal R}=-{\cal U}_0{\cal R}{\cal U}_0^{\dagger}\,,\qquad \tilde{\cal R}_x=
     -{\cal U}_{1}{\cal R}{\cal U}_{1}^{\dagger}
     ={\cal U}_3{\cal R}{\cal U}_3^{\dagger}\,,\qquad
    \tilde{\cal R}_y=-{\cal U}_{2}{\cal R}{\cal U}_{2}^{\dagger}={\cal
    U}_4{\cal R}{\cal U}_4^{\dagger}\,,
\end{equation}
where the explicit form of the $SU(2)$-transformations is
\begin{equation}\begin{array}{ll}
      {\cal U}_0=e^{i\frac{\pi}{2}\tilde{\cal R}_y}=i\tilde{\cal R}_y\,,&\\
     {\cal U}_{1}=e^{i\left(\frac{\pi}{2}{\cal R}-\frac{\pi}{4}\tilde{\cal R}_y\right)}
     =\frac{i}{\sqrt{2}}{\cal R}(1-i\tilde{\cal R}_y)\,,& {\cal U}_3=
     e^{-i\frac{\pi}{4}\tilde{\cal R}_y}
     =\frac{1}{\sqrt{2}}(1-i\tilde{\cal R}_y)\,,\\
    {\cal U}_{2}=e^{-i\frac{\pi}{4}\tilde{\cal R}_x}=
    \frac{1}{\sqrt{2}}(1-i\tilde{R}_x)\,,
    &{\cal U}_4=e^{i\left(\frac{\pi}{2}{\cal R}-
    \frac{\pi}{4}\tilde{\cal R}_x\right)}=
    \frac{i}{\sqrt{2}}{\cal R}(1-i\tilde{\cal R}_x)\,.
\end{array}\label{UU}
\end{equation}
Let us note that the transformations (\ref{RU}) together with
(\ref{RJ}) suggest that the twisted reflections $\tilde{\cal R}_x$
and $\tilde{\cal R}_y$ can be written formally in the following
way\,:
\begin{equation}
    -\tilde{\cal R}_x=\exp (i\pi {\cal U}_3J{\cal U}_3^{\dagger})\,,
    \qquad \tilde{\cal R}_y=\exp (i\pi {\cal U}_1J{\cal U}_1^{\dagger})\,.
\end{equation}

Inspect now how these transformations change the other constituents
of the supersymmetry, Hamiltonian and supercharges. Formally,
Hamiltonian $H_{1/2}$ commutes with any of ${\cal R}$, $\tilde{\cal
R}_{x}$ or $\tilde{\cal R}_y$ so that it is invariant with respect
to the $SU(2)$-transformations (\ref{U}). The formal operator
$\tilde{Q}_{1/2}$ is transformed as
\begin{equation}\begin{array}{lll}
     {\cal U}_{0}\tilde{Q}_{1/2}{\cal U}_{0}^{\dagger}=
    \tilde{Q}_{1/2}\,,&
     {\cal U}_{1}\tilde{Q}_{1/2}{\cal U}_{1}^{\dagger}=
     -\tilde{Q}_{1/2}\,,& {\cal U}_{3}
     \tilde{Q}_{1/2}{\cal U}_{3}^{\dagger}=\tilde{Q}_{1/2}\,,\\
     &{\cal U}_{2}\tilde{Q}_{1/2}{\cal U}_{2}^{\dagger}=
     i\tilde{\cal R}_x\tilde{Q}_{1/2}\,,&{\cal U}_{4}
     \tilde{Q}_{1/2}{\cal U}_{4}^{\dagger}
     =i\tilde{\cal R}_x\tilde{Q}_{1/2}\,.
\end{array}\label{QU}
\end{equation}
Let us suppose that we take the self-adjoint extension
$H_{1/2}^{0}$ (with the supercharge $\tilde{Q}_{1/2}^{0}$) as the
initial system. The transformations (\ref{U}) are unitary, and
applied to $\tilde{Q}_{\alpha}^{0}$ produce self-adjoint
operators, defined on
$\mathcal{U}_k\mathcal{D}(\tilde{Q}_{\alpha}^{0})$ for
$k\in\{0,1,2,3,4\}$. The transformed supercharges in the upper
line of (\ref{QU}) coincide formally with $\tilde{Q}_{1/2}$. As we
found in the previous section, there exists one-parameter family
of the self-adjoint extensions of this operator, labeled by
$\gamma$. Hence, the systems produced by $\mathcal{U}_0$,
$\mathcal{U}_1$ and $\mathcal{U}_3$ should fit into this
classification scheme. This is indeed the case: we can write
\begin{equation}\label{UQgamma}
     {\cal U}_0\tilde{Q}_{1/2}^{0}{\cal U}_0^{\dagger}=
    \tilde{Q}_{1/2}^{\pi},\quad {\cal U}_1\tilde{Q}_{1/2}^{0}
    {\cal U}_1^{\dagger}=\tilde{Q}_{1/2}^{\pi/2},\quad
    {\cal U}_3\tilde{Q}_{1/2}^{0}{\cal U}_3^{\dagger}=
    \tilde{Q}_{1/2}^{3\pi/2}\, ,
\end{equation}
where the value of the index $\gamma$ coherently reflects the domain
of definition, given by (\ref{4-1a}). The remaining systems with the
supercharges of the lower line in (\ref{QU}) do not belong to the
family of self-adjoint operators $\tilde{Q}_{1/2}^{\gamma}$ as
neither of the supercharges coincides formally with
$\tilde{Q}_{1/2}$. The explicit form of the domains of definition of
the new supercharges for $k\in\{1,2,3,4\}$ can be written in the
following compact form\,:
\begin{equation}\label{domu}
 \begin{array}{ccl}
    {\cal U}_k\mathcal{D}\left( \tilde{Q}_{1/2}^0 \right)&=& \left\{
    \Phi(r,\varphi)=f(r,\varphi)+ A\
    K_{\alpha}(r)(1+i^ke^{-i\varphi})\,|\,
    f(r,\varphi)\in \mathcal{D}\left( \overline{Q}_{\alpha} \right),\,
    A\in \mathbb{C} \right\}\,.\end{array}
\end{equation}

Hence, for the semi-integer values of the magnetic flux $\alpha$ we
have a three parametric family of the systems with hidden
supersymmetry, associated with the formal supercharge operator
\begin{equation}\label{UQU}
      \mathcal{Q}_{\alpha}(\beta_x,\beta_y,\beta)=\mathcal{U}
    \tilde{Q}_{\alpha}^0\mathcal{U}^{\dagger},
    \quad  \mathcal{D}(\mathcal{Q}_{\alpha}(\beta_x,\beta_y,\beta))=\mathcal{U}
    \mathcal{D}(\tilde{Q}_{\alpha}^0),\quad \mathcal{U}=
    \mathcal{U}(\beta_x,\beta_y,\beta)\,.
\end{equation}
These systems fit into the general scheme of the self-adjoint
extensions of the Aharonov-Bohm model discussed in
\cite{stovicek}, where the self-adjoint extensions of $H_{\alpha}$ with broken
rotational symmetry were observed. Despite the rotational symmetry is broken in
our present case as well (see (\ref{domu})), domains of definition  are invariant
with respect to the operator $\tilde{J}(\beta_x,\beta_y,\beta)={\cal U}J{\cal
U}^{\dagger}$, i.e. the systems associated with (\ref{UQU}) are unitarily
equivalent to the systems with rotational symmetry.

\section{Superconformal symmetry\label{dynamical symmetries}}

Jackiw showed that like a  charge-monopole system \cite{monopole},
the original Aharonov-Bohm model is characterized by a dynamical
conformal $so(2,1)$ symmetry \cite{jackiw}. We revealed the hidden
$N=2$ supersymmetry not only in the Aharonov-Bohm system
characterized by a regular behavior of the wave functions at the
origin, but also in exotic models corresponding to some special
cases of the $U(2)$ family of self-adjoint extensions of the
formal Hamiltonian operator (\ref{1-3}). On the other hand, if we
look at the $U(2)$ family of the self-adjoint extensions requiring
the scale symmetry, this also excludes immediately  those cases
which are characterized by the presence of the bound states. Such
a similarity with restrictions imposed by the requirement of the
presence of the hidden supersymmetry, certainly, is worth a more
in-depth look. In this section, we study the question of
compatibility of the revealed hidden supersymmetric structure with
the dynamical conformal symmetry.

Besides the Hamiltonian of the system, which we denote here by
$H$, the dynamical conformal symmetry \cite{de Alfaro} is generated by the
operators $D$ and $K$ that depend explicitly on time. They satisfy
equation $ \frac{d}{dt}{\cal C}=\partial_t{\cal C}-i[H,{\cal
C}]=0$, ${\cal C}=D,K$, and their explicit form is given by
\begin{eqnarray}\label{conf}
     D&=&tH-\frac{1}{4}(\vec{x}\vec{\cal{P}}+\vec{\cal{P}}\vec{x})\,
     ,
     \nonumber\\
     K&=&-2t^2H+4tD+\frac{1}{2}\vec{x}^2\, .
\end{eqnarray}
The operator $D$ generates dilatations, while $K$ is the generator
of the special conformal transformations. The conformal algebra
$so(2,1)$ is established by the formal commutation relations
\begin{equation}\label{so(2,1)}
     [D,K]=iK\,,\quad [H,K]=4iD\,,\quad [H,D]=iH\,.
\end{equation}
The  domain $\mathcal{D}_{c}$ where the commutators are well defined
has to be specified. It has to be located in the intersection of the
domains of all the involved operators. Also, the action of each of
the operators $H$, $K$, and $D$ has to keep the wave function in the
domains of the two remaining operators. For $t=0$, the explicit form
of $D$ and $K$ in polar coordinates is
\begin{equation}\label{DKrad}
     D=\frac{i}{2}(1+r\partial_r)\,,
     \quad K=\frac{1}{2}r^2\,.
\end{equation}
Both of these operators are essentially self-adjoint. Indeed, the
solutions of  $Df(r,\varphi)=\pm if(r,\varphi)$ are not square
integrable, while deficiency equations $Kf(r,\varphi)=\pm
if(r,\varphi)$ do not have solutions at all. The domains of  the
essentially self-adjoint operator $K$, and of the self-adjoint
operator $\overline{D}$ are described in Appendix B. Fixing $H$ to
be one of the operators $H_{\alpha}^{AB}$ or
$H_{\alpha}^{\gamma}$, $\gamma=0,\pi$, we can write
\begin{equation}\label{DC}
    \mathcal{D}_{c}=\{\Phi(r,\varphi)\in\mathcal{D}(H)
    \cap\mathcal{D}(\overline{D})\cap\mathcal{D}(K)
    |H\Phi\in\mathcal{D}(\overline{D})\cap\mathcal{D}(K),
    \overline{D}\Phi\in\mathcal{D}(H)\cap\mathcal{D}(K),
    K\Phi\in\mathcal{D}(\overline{D})\}.
\end{equation}
This set is dense in $L_2(\R^2)$ as it contains smooth functions
with compact support ($C_0^{\infty}(\R^2)$).

The generalized eigenvectors (scattering states) of $H$ do not
have compact support, and are not square integrable. However, they
can serve to construct the wave packets which are normalizable,
and represent physical states. These square integrable functions
inherit some of the properties of the scattering states; they do
not belong to $C_0^{\infty}(\R^2)$, and have a specific behavior
of partial waves near the origin, dictated by $\mathcal{D}(H)$. We
can ask whether they are present in $\mathcal{D}_{c}$. The
necessary condition is that the operators $K$ and $D$ do not alter
asymptotic behavior of the partial waves near the origin.

The domain of definition of either $H_{\alpha}^{AB}$ or
$H_{\alpha}^{\gamma}$ is rotationally invariant. The partial waves
near the origin may not be more divergent than a fixed power of $r$,
prescribed by the domain of definition. Keeping in mind explicit
form (\ref{DKrad}), we see that neither $K$ nor $D$ violate this
restriction on asymptotic behavior of partial waves. Hence, the
domain $\mathcal{D}_{c}$ includes the physically interesting
states\footnote{We have in mind two-dimensional exponentially
decreasing (gaussian) wave packet for instance. } composed of the
scattering states. This conclusion is not evident for other
self-adjoint extentions $H_{\alpha}^{\gamma}$ when a general value
of $\gamma$ is considered. Let us just note that the invariance with
respect to $D$ is broken in general. The scale invariance is
recovered for $\gamma=0$ or $\gamma=\pi$ when $\alpha$ is treated as
a free parameter. For fixed  value of the magnetic flux $\alpha=1/2$
${\rm mod}\, 1$, the scale symmetry appears in the whole family of
self-adjoint extensions $H_{1/2}^{\gamma}$ for any value of
$\gamma$, see Appendix B. Therefore, the hidden supersymmetry of the
systems represented by $H_{\alpha}^{AB}$ and $H_{\alpha}^{\gamma}$
comes hand in hand with conformal symmetry and the scale invariance
in particular. Below we show that the both structures are compatible
in the Lie algebraic sense, and give rise to the superconformal
$osp(2|2)$  symmetry.

The operators $K$ and $\overline{D}$ commute with ${\cal R}$, their
domains are invariant with respect to the action of ${\cal R}$, and
they can be  treated as bosonic generators in the framework of the
extended superalgebra. The relevant commutation and anticommutation
relations have to be computed to verify that the superalgebra is
closed. The computation does not depend on the actual choice of the
self-adjoint extension, so that we adopt the notation $H$ for
$H_{\alpha}^{AB}$ or $H_{\alpha}^{\gamma}$, and, respectively,
$Q_1={\cal P}_x+i \varepsilon {\cal  R P}_y$ for $Q_{\alpha}^{AB}$
or $\tilde{Q}_{\alpha}^{\gamma}$, and  $Q_2=-i \varepsilon {\cal R}
Q_1$, where $\varepsilon=+1$ or $-1$ in dependence on the value of
the flux $\alpha$, see Eqs. (\ref{QAB}) and (\ref{Qgamma}). The
self-adjoint generator of dilatations is denoted below by $D$. To
close the superalgebra, two additional integrals of motion
(explicitly dependent on time) have to be involved. In the
commutator of $K$ and $Q_j$, there appear new integrals of motion
\begin{equation}
    [Q_j,K]=-iS_j,\qquad S_2=-i \varepsilon {\cal R}S_1\, ,
\end{equation}
where
$$
    S_1=X+i\varepsilon {\cal R}Y,\quad X=x-2t{\cal P}_x,\quad Y=y-2t{\cal P}_y.
$$
The mixed anticommutator of $Q_j$ and $S_k$ brings a new conserved
quantity, $\{Q_1,S_2\}=2F$,
\begin{equation}\label{FJR}
    F=\varepsilon{\cal R}-J\,.
\end{equation}
Completing the remaining relations dictated by the superalgebra, we
end up with
\begin{eqnarray}
     &\{Q_i,Q_j\}=2\delta_{ij}H\,,\quad \{S_i,S_j\}=4\delta_{ij}K\,,&\nonumber\\
     &\{Q_j,S_k\}=
     -4\delta_{jk}D+2\epsilon_{jk}F\,,&\nonumber\\
     &[Q_j,K]=-iS_j\,,\quad [S_j,K]=0\,,&\nonumber\\
     &[Q_j,D]=\frac{i}{2}Q_j\,,\quad [S_j,D]=-\frac{i}{2}S_j\,,&\nonumber\\
     &[Q_j,H]=0\,,\quad [S_j,H]=2iQ_j\,,&\nonumber\\
     &[F,Q_j]=i\epsilon_{jk}Q_{k}\,,\quad [F,S_j]=i\epsilon_{jk}S_{k}\,,&\nonumber\\
     &[F,H]=[F,K]=[F,D]=0\,,&\nonumber\\
     &[F,{\cal R}]=[H,{\cal R}]=[D,{\cal R}]=
     [K,{\cal R}]=\{Q_j,R\}=\{S_j,{\cal R}\}=0\,.&
     \label{superextend}
\end{eqnarray}
Instead of the even generators $J$ and  ${\cal R}$, in  addition to
the linear combination (\ref{FJR}) we define the operator
\begin{equation}\label{ZJR}
     {\cal Z}=J-\frac{\varepsilon}{2}{\cal R}\,,
\end{equation}
which commutes with all the other even and odd generators of
superalgebra, playing the role of its central charge. The
introduced operators $S_j$, $F$ and ${\cal Z}$ are essentially
self-adjoint on their natural domains of definition, see Appendix
B. Note that from relation $J=F+2{\cal Z}$ it follows that $Q_i$
and $S_i$ are vector operators.

Likewise in the case of the conformal symmetry, the actual domain
of definition $\mathcal{D}_{sc}$ has to be specified to make the
relations (\ref{so(2,1)}) and (\ref{superextend}) consistent. It has to be an intersection of the
domains of the involved operators (just let us remind that
$\mathcal{D}(Q_1)=\mathcal{D}(Q_2)$ and
$\mathcal{D}(S_1)=\mathcal{D}(S_2)$), and the action of any of
them has to keep the function in the intersection of the domains
of the remaining operators.

The same analysis applies as in the case of $\mathcal{D}_{c}$. The
domain $\mathcal{D}_{sc}$ is dense in $L_2(\R^2)$ as it contains
the set of smooth functions with compact support. We require that
neither of the operators violates asymptotics of the functions
near the origin - they should maintain or increase the power of
the leading term in the asymptotic expansion. This requirement is
met by all the new operators $S_j$, $F$ and ${\cal Z}$. Hence, the
domain $\mathcal{D}_{sc}$ can support physically interesting
states represented particularly by wave packets.

We conclude that the three self-adjoint extensions
$H^{AB}_{\alpha}$, $H_{\alpha}^0$ and $H_{\alpha}^{\pi}$ possess the
scale invariance as a consequence of their conformal symmetry. The
conformal and hidden supersymmetric structures of these systems are
compatible, and  lead to the superconformal symmetry. The resulting
algebraic structure corresponds to the superalgebra $osp(1|2)\times
o(2)$, which was observed earlier in various physical models
\cite{superconf,LebLoz,supernon,superconformal}, including a
spin-$1/2$ particle in the presence of a magnetic vortex
\cite{superconformal}~\footnote{The analysis of the algebraic
structure was performed in \cite{superconformal} on a formal level,
without touching the questions of self-adjointness of corresponding
generators.}. For spin-$1/2$ particle systems possessing the
superconformal symmetry, the role of the grading operator is played
by the matrix $\sigma_3$. We revealed here the same superalgebraic
structure in the system without fermionic degrees of freedom.

\section{Hidden supersymmetry and anyons\label{sectionanyons}}

In early eighties, Wilczek  proposed a dynamical mechanism for
realization of anyons  that is based on the Aharonov-Bohm effect
\cite{wilczek}. Here we show that the anyon picture  provides a
rather natural interpretation for the hidden supersymmetric
structure described in the previous sections.

Consider a two-anyon, planar  system
described by the formal
Hamiltonian operator
\begin{equation}\label{Hany}
    H_{any}=
   2\sum_{I=1}^2\left(\vec{p}_I-\vec{a}_I(\vec{r})
\right)^2,
\end{equation}
where $\vec{p}_I=-i\partial/\partial \vec{x}_I$, $\vec{r}$ is a
relative coordinate, $\vec{r}=\vec{x}_1-\vec{x}_2$, and we set the
masses of the constituents $m_1=m_2=4$. The constituent point
particles are `statistically charged', and each carries a `magnetic'
vortex described by the statistical vector potential,
\begin{equation}\label{anyA}
    a_1^k(\vec{r})= - a_2^k(\vec{r})=
    \frac{1}{2}\alpha
    \epsilon^{kl}\frac{r^l}{\vec{r}\,{}{}^2}\,.
\end{equation}
In the center of mass reference frame, Hamiltonian (\ref{Hany})
takes the form (\ref{1-3}).

The two-anyon system may be composed from statistically interacting
identical bosons, or fermions. This means that when the statistical
interaction is switched off $(\alpha=0$), the constituent particles
should obey either Bose, or Fermi statistics. The identity of the
particles and their statistics are taken into account by requiring
that under the exchange, $\vec{r}\rightarrow -\vec{r} \Rightarrow
\varphi\rightarrow \varphi+\pi$,  the wave function obeys the
relation
$\psi_{\alpha=0}(r,\varphi+\pi)=\eta\psi_{\alpha=0}(r,\varphi)$ with
$\eta=+1$ for boson, or $\eta=-1$ for  fermion constituents.  This
requirement is maintained when the statistical interaction is
switched on. Therefore, we have
\begin{equation}\label{nature}
     \psi_{\alpha}(r,\varphi)=\sum_le^{il\varphi}f_{\alpha,l}(r),
     \quad l\in\left\{\begin{array}{l}2\mathbb{Z}\ \mbox{for anyons based on bosons\,,}\\
    2\mathbb{Z}+1\ \mbox{for anyons based on fermions\,.}
    \end{array}\right.
\end{equation}
Requiring the Hamiltonian (\ref{Hany}) to be self-adjoint operator,
its domain of definition has to be specified. The nontrivial
behavior of the wave functions  (\ref{nature}) near the origin,
dictated by the particular choice of the self-adjoint extension, is
interpreted then as a contact (zero-range) interaction of the anyons
\cite{manueltarrach}.

The anyon framework provides an interesting interpretation for the
systems we studied in Sections \ref{ABSUSY} and \ref{exotic
models}. As follows from the discussion at the end of Section
\ref{ABSUSY}, Hamiltonians $H_{\alpha}^{AB}$, $H_{\alpha}^{0}$ and
$H_{\alpha}^{\pi}$ can be described by diagonal two-by-two
matrices in representation where the grading operator
$\mathcal{R}$  is given by the  Pauli matrix $\sigma_3$, see
(\ref{matrixham}). The upper and lower components of the states in
this representation correspond to $\pi$-periodic and
$\pi$-antiperiodic in $\varphi$ parts of the wave functions. Due
to the described correspondence between the Hamiltonian
(\ref{Hany}) of the two-anyon system and the Hamiltonian
(\ref{1-3}), we conclude that the diagonal components
$\Pi_{\pm}H_{\alpha}^{AB}$ and $\Pi_{\pm}H_{\alpha}^{\gamma}$ of
the studied spinless systems can be understood as the Hamiltonians
that describe the relative motion of the two-anyon systems. The
upper component represents the system based on bosons (as it acts
on $\pi$-periodic functions) while the lower-diagonal operators
rule the dynamics of the system based on fermions. The
self-adjoint Hamiltonians $\Pi_{\pm}H_{\alpha}^{\gamma}$ imply
additional contact interaction of the anyons.

Therefore, the \emph{hidden} superconformal symmetry that we
revealed in spinless Aharonov-Bohm system corresponds to an
\emph{explicit} center-of-mass supersymmetric structure  of the
system composed from the \emph{two } two-anyon systems based on
bosons \emph{and} on fermions.

\vspace*{4mm}

\section{Discussion and outlook\label{outlook}}

Let us summarize and discuss  the main results by stressing
the  physical aspects  that are behind the revealed hidden
supersymmetric structure.

For integer values of the flux, $\alpha=n,$ $n\in \Z$, the
Aharonov-Bohm system   is unitary equivalent to a planar
free particle system ($\alpha=0$). The latter possesses the
rotation and translation symmetries generated by the
angular momentum operator $J$, and by mutually commuting
momenta operators ${\cal P}_x$ and ${\cal P}_y$. In
correspondence with this, Hamiltonian operator (\ref{1-3})
can be factorized as
\begin{equation}
    H_n=({\cal P}_x+i\epsilon{\cal P}_y)({\cal
    P}_x-i\epsilon{\cal P}_y), \label{H0f}
\end{equation}
 or, alternatively,
can be presented as a perfect square,
\begin{equation}
    H_n=({\cal P}_x+i\epsilon {\cal R}{\cal P}_y)^2,
\label{H0Q}
\end{equation}
where the parameter $\epsilon$ can take any of two values, $+1$ or
$-1$, and ${\cal R}$ is a nonlocal operator of rotation for angle
$\pi$. For $\alpha\neq n$, the formal Aharonov-Bohm Hamiltonian
(\ref{1-3}) can also be factorized in the form (\ref{H0f}), or
(\ref{H0Q}). However, in the case of non-integer flux, the
operators ${\cal P}_x$ and ${\cal P}_y$ are not physical, and the
translation invariance is broken, see below. Thus, for $\alpha\neq
n$, (\ref{H0f}) is a purely \emph{formal} factorization. In
contrast with (\ref{H0f}), representation (\ref{H0Q}) can be well
defined. A nontrivial property associated with factorization
(\ref{H0Q}) is that for a given flux $\alpha\neq n$, two different
choices for the value of the parameter $\epsilon$ correspond to
physically distinct systems. For $\alpha \in (-1,0)$ ${\rm mod}$
$2$, $\epsilon=+1$, and $\alpha \in (0,1)$ ${\rm mod}$ $2$,
$\epsilon=-1$, factorization (\ref{H0Q}) corresponds to the
original system $H_\alpha^{AB}$ investigated by Aharonov and Bohm
\cite{aharonov,AB2}, which is characterized by a regular at the
origin  behavior of the Hamiltonian eigenfunctions. Alternative
choice of the values of the parameter $\epsilon$ in (\ref{H0Q})
gives rise to two different, \emph{exotic} models given by
self-adjoint Hamiltonians $H^\gamma_\alpha$ with $\gamma=0,\,
\pi$, which are characterized by a singular behavior at the origin
of their eigenfunctions in one specific partial wave correlated
with the value of the flux, see Eqs. (\ref{f1})--(\ref{f4}). For
half-integer values $\alpha=n+1/2$, both exotic systems with
$\gamma=0$ and $\gamma=\pi$ are unitary equivalent, and like the
Aharonov-Bohm model  $H_\alpha^{AB}$, they possess additional
nonlocal integrals of motion in the form of the twisted reflection
operators $\tilde{\cal R}_x$ and $\tilde{\cal R}_y$.  These
nonlocal integrals together with ${\cal R}$ satisfy the same
algebraic relations as the three Pauli matrices, i.e. generate a
spinorial representation of $su(2)$ realized on the states of the
corresponding system.

Identifying the nonlocal operator ${\cal R}$ as the $\Z_2$-grading
operator, we interpret the self-adjoint operator appearing in
factorization (\ref{H0Q}) as the supercharge $Q_1$, another
self-adjoint supercharge is $Q_2=i{\cal R}Q_1$. Therefore, for
non-integer flux values, the translation symmetry of the
Aharonov-Bohm system $H_n^{AB}$ is broken, and corresponding
mutually commuting generators ${\cal P}_x$ and ${\cal P}_y$ are
substituted by nonlocal, mutually anti-commuting, odd operators
$Q_1={\cal P}_x+i\epsilon {\cal R}{\cal P}_y$ and $
Q_2=-\epsilon{\cal P}_y+i {\cal R}{\cal P}_x$ \footnote{This
picture can be compared  loosely with that appearing in the
BRST-scheme of quantization of usual, non-supersymmetric gauge
invariant theories, where after gauge fixing the even generators
of gauge symmetries are substituted by the mutually anti-commuting
nilpotent BRST and anti-BRST operators \cite{BRST}.}.

By taking into account the dynamical conformal symmetry, the
revealed hidden supersymmetric structure of the spinless
Aharonov-Bohm system is extended to the superconformal $osp(2\vert
2)$ symmetry. By this superconformal symmetry, one can relate not
only the states with the same value of the angular momentum and
different values of the energy, see Ref. \cite{jackiw}, but also
the states with different energy values and different in one
angular momentum  in correspondence with Fig. \ref{susyex}
\cite{CFJP_prep}.

We have shown that the  \emph{hidden} superconformal symmetry of
the spinless Aharonov-Bohm system is in one-to-one correspondence
with \emph{explicit} center-of mass supersymmetric structure of
the system composed from the two two-anyon  subsystems, the
composites of one of which before switching on statistical
interaction ($\alpha=0$) satisfy boson statistics, while another
subsystem is formed by two identical fermion particles. The exotic
models given by the Hamiltonians $H_\alpha^\gamma$, $\gamma=0,\,
\pi$, with nontrivial behavior of the wave functions near the
origin correspond in this interpretation to the case of anyons
with a contact (zero-range) interaction.

The hidden supersymmetric structure is reflected in the scattering
picture. To see this, consider the case of the Aharonov-Bohm model
given by the Hamiltonian $H_\alpha^{AB}$. Its regular at the
origin eigenfunctions, which  correspond to a plane wave incident
from the right ($x=+\infty$, $y=0$), have a form
\cite{aharonov,Ruij,HagenPRD}
\begin{equation}
    \psi=\sum_{l=-\infty}^{\infty}a_l e^{il\varphi}{\cal
    J}_{\vert l+\alpha\vert}(kr),
    \label{psiscat}
\end{equation}
$H_\alpha^{AB}\psi=k^2\psi$, where
\begin{equation}
    a_l=e^{-i\frac{\pi}{2}\vert l+\alpha\vert}.
    \label{al}
\end{equation}
For the sake of definiteness, suppose that $\alpha\in
(-1,0)$. In this case, the coefficients (\ref{al}) satisfy
the relation
\begin{equation}\label{a2l}
    a_{2l}=\varepsilon ia_{2l-1}, \quad
    {\rm where}\quad \varepsilon=\left\{\begin{array}{cll}
-1&
\mbox{for}& l\geq 1,\\
+1& \mbox{for}& l\leq 0 .
\end{array}\right.
\end{equation}
Acting on (\ref{psiscat}) by the supercharge (\ref{2-3}),
and taking into account the reccurance relations satisfied
by the Bessel functions,
$$
    {\cal J}_{\nu\mp 1}(x)=\left(\pm
    \frac{d}{dx}+\frac{\nu}{x}\right)
{\cal J}_\nu(x),
$$
and relation (\ref{a2l}), we find that the energy
eigenfunctions (\ref{psiscat}) are simultaneously  the
supercharge eigenstates, $Q_\alpha^{AB}\psi=-k\psi$. The
second supercharge (as well as the operator ${\cal R}$)
transforms the state (\ref{psiscat}) into another
eigenstate of $H_\alpha^{AB}$, which corresponds to the
plane wave incident from the left.

Making use of relation (\ref{a2l}), energy eigenfunction
(\ref{psiscat}) can be presented  as a superposition of the
supercharge  eigenstates (\ref{lnozeroscat}),
\begin{equation}\label{QHeigen}
    \psi=\sum_{l=-\infty}^{0}\Phi_l^-
    +\sum_{l=1}^{+\infty}
    \Phi_l^+,
\end{equation}
where
\begin{equation}\label{QH-}
    \Phi_l^-(r,\varphi)=e^{i\frac{\pi}{2}\alpha}
    (-1)^le^{2il\varphi}\left({\cal
    J}_{-(2l+\alpha)}(kr)
    -ie^{-i\varphi}{\cal J}_{-(2l-1+\alpha)}(kr)\right),
\end{equation}
\begin{equation}\label{QH+}
    \Phi_l^+(r,\varphi)=e^{-i\frac{\pi}{2}\alpha}
    (-1)^le^{2il\varphi}\left({\cal
    J}_{2l+\alpha}(kr)
    +ie^{-i\varphi}{\cal J}_{2l-1+\alpha}(kr)\right),
\end{equation}
$Q_\alpha^{AB}\Phi_l^-=-k\Phi_l^-$, $l=0,-1,-2,\ldots$,
$Q_\alpha^{AB}\Phi_l^+=-k\Phi_l^+$, $l=1,2,\ldots$. The
energy eigenstate ${\cal
R}\psi(r,\varphi)=\psi(r,\varphi+\pi)$, that corresponds to
the plane wave incident from the left, is the eigenstate of
the supercharge of the eigenvalue $+k$,
$Q_\alpha^{AB}\psi(r,\varphi+\pi)=+k\psi(r,\varphi+\pi)$.
The superpositions $\psi(r,\varphi)\pm i
\psi(r,\varphi+\pi)$ are the eigenstates of the second
supercharge $Q_2=i{\cal R}Q_\alpha^{AB}$,
$Q_2(\psi(r,\varphi)\pm i \psi(r,\varphi+\pi))=\mp
k(\psi(r,\varphi)\pm i \psi(r,\varphi+\pi))$. For
$\alpha=-1/2$, the states (\ref{QHeigen}) and ${\cal
R}\psi$ form the invariant subspace also for two additional
nonlocal integrals of motion that appear in the system in
this case, $\tilde{\cal R}_x=-ie^{i\varphi}{\cal R}_x$,
$\tilde{\cal R}_y=e^{i\varphi}{\cal R}_y$, where ${\cal
R}_x:\, \varphi\rightarrow \pi-\varphi$, ${\cal R}_y:\,
\varphi\rightarrow -\varphi$.

\vskip0.2cm

The non-physical nature of the operators ${\cal P}_x$ and ${\cal
P}_y$ can be revealed immediately if to  apply them to the
Hamiltonian eigenfunction (\ref{psiscat}). The action of the
operator ${\cal P}_x+i{\cal P}_y$ produces a state, in which the
$l=1$ partial wave is multiplied by the function ${\cal J}_{\vert
\alpha\vert -1}(kr)$, that has a not permitted, singular behavior
at the origin. Analogously, the state $({\cal P}_x-i{\cal
P}_y)\psi$ contains a partial wave with $l=0$  multiplied by the
singular at the origin function ${\cal
J}_{-\vert\alpha\vert}(kr)$. The supercharge (\ref{2-3}) can be
written in the form $Q_\alpha^{AB}=\Pi_+({\cal P}_x+i{\cal
P}_y)+\Pi_- ({\cal P}_x-i{\cal P}_y)$. Its projectors  on the
subspaces with even and odd $l$, $\Pi_\pm=\frac{1}{2}(1\pm{\cal
R})$, just annul the singularities produced by nonphysical
operators ${\cal P}_x\pm i{\cal P}_y$ in corresponding partial
waves. One can show that in the case of the exotic systems
$H^\gamma_\alpha$, $\gamma=0,\,\pi$, considered in Section
\ref{exotic models}, the picture is similar: the operators ${\cal
P}_x\pm i{\cal P}_y$ acting on the states of the domain of the
Hamiltonian $H^\gamma_\alpha$, in contrast with the action of the
supercharges, produce the states that do not belong to the domain.
This explains the mechanism of translation symmetry breaking, and
its substitution for the hidden supersymmetry, as well as a purely
formal character of factorization (\ref{H0f}).
 Note also here that in the case $\alpha=n$, the
action of the operators ${\cal P}_x\pm i{\cal P}_y$ on the energy
eigenstates (\ref{psiscat}) does not produce singularities, and
operators ${\cal P}_x$ and ${\cal P}_y$ commute on the domain of
the Hamiltonian $H_n^{AB}$. This corresponds to a unitary
equivalence of the model  $H_n^{AB}$ to a free planar particle
system discussed in Section \ref{ABSUSY}.

Partial wave analysis applied to the wave function
(\ref{psiscat}) gives the scattered wave with asymptotic
behavior for large $r$, see \cite{aharonov,HagenPRD},
$\psi_{sc}\rightarrow r^{-1/2}e^{ikr}f(\varphi)$ ,
$$
    f(\varphi)=(2\pi i k)^{-1/2}\sum_{l=-\infty}^{+\infty}
    e^{il(\varphi-\pi)}\left(e^{2i\delta_l}-1\right),
$$
where the phase shifts are given by
\begin{equation}\label{phaseshift}
    \delta_l=-\frac{\pi}{2}\vert l+\alpha \vert
    +\frac{\pi}{2}\vert l\vert\,.
\end{equation}
With taking into account (\ref{al}) and (\ref{a2l}), we get
the relation
\begin{equation}
    e^{2i\delta_{2l}}=e^{2i\delta_{2l-1}}.
    \label{delta2l}
\end{equation}
This relation between the phase shifts reflects coherently
with the picture presented on Fig. \ref{susyex} a hidden
supersymmetry in the scattering problem of the spinless
Aharonov-Bohm model in the case $\alpha \in (-1,0)$ mod
$2$. In the case $\alpha \in (0,1)$ mod $2$, index $2l-1$
on the right hand side of relation (\ref{delta2l}) is
changed for $2l+1$ in correspondence with Fig.
\ref{susyex1}.

Finally, we note that the original Aharonov-Bohm calculation of
the scattering amplitude \cite{aharonov}, mathematically more
justified in comparison with partial wave analysis, see
\cite{HagenPRD}, was based on separation of the wave function
(\ref{psiscat}) into three functions, $\psi=\psi_1+\psi_2+\psi_3$.
In the case $\alpha\in (-1,0)$, this corresponds to separation of
wave function $\psi$ in partial wave with $l=0$ ($\psi_3$), and in
the infinite sums with $l>0$ ($\psi_1$) and $l<0$ ($\psi_2$)
\cite{aharonov,HagenPRD}. For the function $\psi_1$ the equivalent
integral representation was found in \cite{aharonov}, that allowed
the authors to find its asymptotic expansion, and then to
calculate the scattering amplitude. The function $\psi_1$  is
nothing else as the second series  in (\ref{QHeigen}). This means that
the original method  used in \cite{aharonov} is coherent with the
hidden supersymmetric structure revealed in the present paper.

\vskip 0.2cm In conclusion, let us discuss some open problems to
be interesting for further investigation. \vskip 0.1cm

The stationary Schr\"odinger equation of the Aharonov-Bohm model
is separable in polar coordinates. Its radial equation corresponds
to stationary Schr\"odinger equation of Calogero model. When we
specify the self-adjoint extension of the formal Hamiltonian
operator $H_{\alpha}$, the self-adjoint extension of the radial
part of $H_{\alpha}$ is fixed as well. In other words, fixing the
value of the angular momentum, the (rotationally invariant)
self-adjoint extension of $H_{\alpha}$ fixes the self-adjoint
extensions of Calogero model \cite{FPconformal}. In \cite{gitman},
Gitman et al. discussed recently the dilatation symmetry of the
self-adjoint extensions of this one-dimensional system. They
concluded that there are only few self-adjoint extensions of the
Calogero model which possess scale invariance. We described three
Aharonov-Bohm type systems, represented by $H_{\alpha}^{AB}$ and
$H_{\alpha}^{\gamma}$, $\gamma=0,\pi$. These systems proved to be
scale invariant. It is quite intriguing question, whether these
two distinct symmetries, scale invariance and hidden
supersymmetry, are interrelated somehow. We suppose that this is
indeed the case. Verification of this hypothesis could provide a
deeper insight into the physical system and its symmetries as
well.

Recently,  the hidden supersymmetry of the reflectionless
P\"oschl-Teller system was explained in \cite{ABAdS} in the
context of non-relativistic  AdS/CFT correspondence
\cite{LeiPl,ADS/CFT}. The rather natural  question is then whether
some AdS/CFT holography interpretation exists for the hidden
superconformal symmetry observed here.

The Aharonov-Bohm type systems described formally by $H_{\alpha}$,
can have up to two bound states. The systems with negative
energies were disqualified in our framework from the very
beginning by requirement of the presence of a self-adjoint
supercharge. This is in correlation with spontaneous breakdown of
their scale invariance. However, such systems could fit into the
framework of the nonlinear supersymmetry. Analysis of this
possibility requires a separate consideration.

We analyzed the spinless particle case. It would be interesting to
consider the systems with spin degrees of freedom as well
\cite{CFJP_prep}. The spin one-half system would be governed by
the Pauli Hamiltonian, whose diagonal components would differ
formally just in the sign of the magnetic field, cf.
\cite{AhCash,superconformal}. This suggests that the actual
self-adjoint extensions of the upper- and the lower-diagonal
elements of the matrix Hamiltonian could differ in some way. The
standard supersymmetry should be present then in addition to the
hidden supersymmetry, at least in some particular cases. The
presence of both, explicit and hidden, supersymmetries should give
rise to the structure of tri-supersymmetry \cite{finitegap,
trisusydelta,CFJP_prep}.

As we observed in Section \ref{semi-integer}, in the case of
half-integer flux values there exists a three parameter family of
unitary transformations (\ref{U}), generated by ${\cal R}$,
$\tilde{\cal R}_x$ and $\tilde{\cal R}_y$. These transformations
do not change formal Hamiltonian $H_{\alpha}$, but interchange its
self-adjoint extensions. Hence, there exists a three parametric
family of self-adjoint extensions of $H_{\alpha}$ which allow the
existence of the hidden supersymmetry, see (\ref{UQU}). We
discussed few particular cases in (\ref{UQgamma}), where the
systems associated with $\tilde{Q}_{1/2}^{\gamma}$ for
$\gamma\in\{0,\pi/2,\pi,3\pi/2\}$ were interrelated by these
unitary mappings. The family  of all the self-adjoint extensions
of $H_{\alpha}$ is four parametric \cite{stovicek}. So it seems
that a great part of the self-adjoint extensions of $H_{\alpha}$
possesses hidden supersymmetry for semi-integer values of
$\alpha$. It would be interesting to clarify this point.

We investigated the question of the presence of the hidden
supersymmetry in spinless quantum mechanical Aharonov-Bohm type
systems. The intriguing open question is whether such a symmetry may
be present in related field systems. The simplest system for such a
generalization could be a non-relativistic (2+1)-dimensional model
of a boson field minimally coupled to a Chern-Simons field
\cite{ChSJ, Bergman, ChSH}. If the hidden bosonized supersymmetry of
the nature discussed here is present in such a field system, then
its  supersymmetrically extended (by inclusion of a fermion field)
version \cite{LebLoz} would be described by a more reach than the
$osp(2|2)$ superconformal structure \cite{supernon}, related to the
tri-supersymmetry \cite{finitegap, trisusydelta}.

\vskip0.3cm \textbf{Acknowledgements} \vskip0.1cm
 \noindent The
work has been partially supported by DICYT (USACH), MECESUP
Project FSM0605, CONICYT  and FONDECYT Grants 1095027, 3085013 and 3100123, Chile, by
the MSMT ÒDoppler
InstituteÓ project LC06002,
and by CONICET (PIP 01787) and UNLP
(Proy.~11/X492), Argentina. H.F. is indebted to the Physics
Department of Santiago University (Chile) for hospitality.

The Centro de Estudios Cient\'{\i}ficos (CECS) is funded by the Chilean Government through the Millennium Science Initiative and the
Centers of Excellence Base Financing Program of Conicyt. CECS is also supported by a group of private companies which at present includes Antofagasta Minerals, Arauco, Empresas CMPC, Indura, Naviera Ultragas and Telef\'onica del Sur.

\setcounter{equation}{0}
\renewcommand{\theequation}{A.\arabic{equation}}

\section*{Appendix A}
Let us present in more detail the procedure of self-adjoint
extension of the following operator
$$
     \hat{Q}={\cal P}_x+i{\cal R}{\cal P}_y\,.
$$
The supercharge (\ref{QAB}) coincides with this operator for
$\alpha\in[-1,0]\ \mbox{mod}\ 2$. $\hat{Q}$ can be identified with
(\ref{Qgamma}) for  $\alpha\in(0,1)\ \mbox{mod}\ 2$ as well.
Hence, the analysis of self-adjoint extensions of $\hat{Q}$ for
any value of the flux will provide, using the unitary
transformation $U_1$ sequently, a complete information on
self-adjoint extensions of both (\ref{QAB}) and (\ref{Qgamma}).

The symmetric operator $Q$ is a restriction of $\hat{Q}$ to
$C_0^{\infty}(\mathbb{R}^2-\{0\})$. The following relation will be
useful in the forthcoming analysis\,:
\begin{equation}\label{2-6}
    \left( \phi, \hat{Q} \psi \right)
    -\left(\hat{Q} \phi, \psi \right)
    = \lim_{r \rightarrow 0^+} \int_0^{2\pi} d\varphi\, r \left[\left(
    -i \cos\varphi +R \sin \varphi\right)\phi(r,\varphi) \right]^*
    \psi(r,\varphi)\,.
\end{equation}
One can easily see that $Q$ is symmetric, since the right hand
side of Eq.\ (\ref{2-6}) vanishes for all $\phi,\psi\in
\mathcal{D}(Q)$.

\textit{\textbf{The adjoint}}.  The adjoint of ${Q}$, $Q^\dagger$,
is a linear operator defined on the set of those functions for which
$\left( \phi, \hat{Q} \psi \right)$ is a linear continuous
functional of $\psi\in\mathcal{D}(Q)$ (see \cite{R-S}, for example).
This requires that for any $\phi \in \mathcal{D}(Q^\dagger)$ there
is a vector $\chi \in \mathbf{L}_2\left( \mathbb{R}^2 \right)$ such
that
\begin{equation}\label{2-7}
    \left( \phi, {Q} \psi \right) =
    \left( \chi, \psi \right)\,, \
    \forall \, \psi \in \mathcal{D}(Q)\,.
\end{equation}
For each $\phi$, this vector is unique (since $\mathcal{D}(Q)$ is
dense in $\mathbf{L}_2\left( \mathbb{R}^2 \right)$) and the action
of the adjoint operator is defined as $Q^\dagger \phi := \chi$.

Since functions $\psi(r,\varphi)\in \mathcal{D}(Q)$ identically
vanish in some neighborhood of the origin, the right hand side of
Eq.\ (\ref{2-6}) vanishes for any function $\phi(r,\varphi)$ such
that $\hat{Q}\phi(r,\varphi)\in \mathbf{L}_2\left( \mathbb{R}^2
\right)$. Therefore, the adjoint operator is densely defined in
\begin{equation}\label{2-8}
    \mathcal{D}\left( Q^\dagger \right) = \left\{
    \phi(r,\varphi)\in \mathcal{AC}
    \left( \mathbb{R}^2 \backslash \left\{ \mathbf{0} \right\}\right)
    \cap \mathbf{L}_2\left( \mathbb{R}^2 \right) :
    \hat{Q}\phi(r,\varphi)\in \mathbf{L}_2\left( \mathbb{R}^2 \right)
    \right\}\,,
\end{equation}
where $\mathcal{AC}
    \left( \mathbb{R}^2 \backslash \left\{ \mathbf{0} \right\}\right)$
is a set of absolutely continuos functions in punctured plane \cite{R-S}.

Since the set $\left\{{e^{i m \varphi}}\,, m\in \mathbb{Z}
\right\}$ is a complete orthogonal system in $\mathbf{L}_2(S^1)$,
we can write
\begin{equation}\label{2-9}
    \phi(r,\varphi)=\sum_{m\in \mathbb{Z}}
    e^{i m \varphi} \, \phi_m(r)\,,
\end{equation}
where $\phi_m(r)\in \mathcal{AC}\left(\mathbb{R}^+\backslash
\left\{0 \right\}\right) \cap \mathbf{L}_2\left(\mathbb{R}^+; r\,
dr \right)$. Then, the condition $\hat{Q}\phi(r,\varphi)\in
\mathbf{L}_2\left( \mathbb{R}^2 \right) $ for $\alpha\notin
\mathbb{Z}$ reduces to
\begin{equation}\label{a-3}
    \left| \phi_{2l}(r) \right| = \left\{
    \begin{array}{c}
      O(1)\,, \quad {\rm for} \  2l+\alpha \notin (0,1)\,,
      \\ \\
      O\left(r^{-(2l+\alpha)}\right)\,, \quad {\rm for} \ 2l+\alpha \in
      (0,1)\,,
    \end{array}
     \right.
\end{equation}
and
\begin{equation}\label{a-4}
    \left| \phi_{2l-1}(r) \right| = \left\{
    \begin{array}{c}
      O(1)\,, \quad {\rm for} \  2l-1+\alpha \notin (-1,0)\,,
      \\ \\
      O\left(r^{(2l-1+\alpha)}\right)\,, \quad {\rm for}
      \ 2l-1+\alpha \in (-1,0)\,.
    \end{array}
     \right.
\end{equation}

For $\alpha=\beta-2l_0\in\mathbb{Z}$, $\beta\in \{0,1\}$, the
partial waves $\phi_j$  are subject to the following restrictions
\begin{equation}\label{B-4}
     |\phi_j|=O(1)\ \mbox{for}\ j\neq 2l_0-\beta\,,\qquad
     |\phi_{2l_0-\beta}|
    = O(\sqrt{-\log \mu r})\,.
\end{equation}

\textit{\textbf{The closure $\overline{Q}$}}. The minimal closed
extension of $Q$ is called the \emph{closure} of this operator,
which is defined as ${\overline{Q}}:= \left( Q^\dagger
\right)^\dagger$. According to the previous discussion on the
definition of the adjoint operator and Eq. (\ref{2-6}), it follows
that its domain is the set of functions $f(r,\varphi)$ for which
$\hat{Q} f(r,\varphi) \in \mathbf{L}_2\left( \mathbb{R}^2 \right)$
and (see (\ref{2-6}))
\begin{equation}\label{2-12}
    \lim_{r \rightarrow 0^+} \int_0^{2\pi} d\varphi\, r \left[\left(
    - i \cos\varphi +R \sin \varphi\right)f(r,\varphi) \right]
    {\phi(r,\varphi)}^* =0\,,\quad
    \forall \, \phi(r,\varphi) \in \mathcal{D}(Q^\dagger)\,.
\end{equation}
To get an insight into the restrictions on $f(r,\varphi)$ posed by
this requirement,  it is convenient to employ the Fourier series of
$f(r,\varphi)$,
\begin{equation}\label{2-14}
    f(r,\varphi)=\sum_{m\in \mathbb{Z}}  e^{i m \varphi} \, f_m(r)\,,\quad
    f_m(r)\in \mathcal{AC}\left(\mathbb{R}^+\backslash \left\{0 \right\}\right)
    \cap \mathbf{L}_2\left(\mathbb{R}^+; r\, dr \right)\,.
\end{equation}
\noindent For $\alpha\notin(0,1)|_{\mod\ 2}$, the conditions posed
on $f_m$ are identical with (\ref{a-3}) and (\ref{a-4}) (resp.
(\ref{B-4})). This means that the domains of definition
$Q^{\dagger}$ and $\overline{Q}$ are identical and the operator $Q$
is essentially self-adjoint\footnote{A densely defined symmetric
operator $A$ is essentially self-adjoint if $\overline{A}=
A^\dagger$.} in this case. Having in mind the note in the begining
of the Appendix, we conclude that the operator $Q_{\alpha}$ defined
in (\ref{QAB}) has unique self-adjoint extension $Q_{\alpha}^{AB}$
for any value of the flux. Its domain of definition can be written
as
\begin{equation}\label{DQAB}
     \mathcal{D}(Q_{\alpha}^{AB})=\left\{f(r,\varphi)=\sum_lf_l(r)
     e^{il\varphi}, f_l\in\mathcal{AC}(\mathbb{R}^+\setminus\{0\})
     \cap L_2(\mathbb{R}^+l;rdr),
     |f_l(r)|= O(1)\right\}\ \mbox{for}\ \alpha\notin\mathbb{Z},
\end{equation}
and for $\alpha=-2l_0+\beta\in\mathbb{Z}$,
\begin{eqnarray}\label{DQABZ}
     \mathcal{D}(Q_{-2l_0+\beta}^{AB})
     &=&\left\{f(r,\varphi)=\sum_lf_l(r)e^{il\varphi},
     f_l\in\mathcal{AC}(\mathbb{R}^+\setminus\{0\})
     \cap L_2(\mathbb{R}^+l;rdr),\right.\nonumber\\
    &&\left.|f_l(r)|= O(1)\ \mbox{for}\ m\neq -2l_0+\beta,
    |f_{2l_0-\beta}|= O(\sqrt{-\log r})\right\}.
\end{eqnarray}

For $2l_0+\alpha\in(0,1)$, the conditions on $f_{2l_0}$ and
$f_{2l_0-1}$ are more restrictive,
\begin{equation}\label{9-12}    \begin{array}{c}
      f_{2l_0}(r)= o\left( r^{-(2l_0+\alpha)} \right)\,,\qquad
      f_{2l_0-1}(r)= o\left( r^{(2l_0-1+\alpha)} \right)\,.
    \end{array}
\end{equation}
This means that the restriction of $Q^\dagger$ to the subspace
$\mathcal{H}_{l_0}$ has a larger domain than the restriction of
$\overline{Q}$ to this subspace. Since these domains do not
coincide, $Q$ is not essentially self-adjoint. Let us remind that
for these values of $\alpha$, $Q$ corresponds to
$\tilde{Q}_{\alpha}$, see (\ref{Qgamma}).

\textit{\textbf{Deficiency subspaces}}. We shall find solutions
$\phi=\phi_{2l}e^{2li\varphi}+\phi_{2l-1}e^{(2l-1)i\varphi}$ of $
Q^{\dagger}\phi=\pm i \mu \phi$ for
$\phi\in\mathcal{D}(Q^{\dagger})$. We can use directly the
equation (\ref{3-1}) for $\lambda=\pm i \mu$. It reduces to
\begin{equation}\label{3-3}
    \phi_{2l}''(r) + \frac{1}{r}\,\phi_{2l}'(r)-
    \left\{ \mu^{2} + \frac{(2l+\alpha)^{2}}{r^{2}}
    \right\} \phi_{2l}(r) =0\,.
\end{equation}
This differential equation has solutions of the form $\phi_{2l}(r) =
C_1 K_{\left|2l+\alpha \right|}(\mu r)+C_2 I_{\left|2l+\alpha
\right|}(\mu r)$, where $I_\nu$ and $K_\nu$ are the modified Bessel
functions of the first and second (or, Macdonald function) kinds,
respectively. The modified Bessel function of the first kind
($I_\nu$ ) has to be discarded as it diverges for $f\rightarrow
+\infty$, $C_2=0$ for all $l$. The function $K_\nu$ decreases
exponentially in infinity. For $r\rightarrow +0$, it reads
\begin{equation}\label{3-5}
    K_\nu(z) \sim 2^{|\nu|-1} \Gamma(|\nu|) \, z^{-|\nu|}
\left(1+O\left(z^2\right) \right)\,.
\end{equation}

We require the eigenvectors of $Q^{\dagger}$ to lie in
$\mathcal{D}(Q^{\dagger})$ and to be square integrable in
particular. This requirement is met only for $0< 2l+\alpha < 1$,
i.e. for $l=l_0$. Then there is one (and only one) eigenvector of
$Q^\dagger$ corresponding to each of the eigenvalues $\lambda = \pm
i \mu$, given by
\begin{equation}\label{3-8}
    \Phi_\pm = e^{i 2 l_0 \varphi} K_{2l_0+\alpha}(\mu r)
    \pm e^{i (2 l_0-1) \varphi} K_{1-(2l_0+\alpha)}(\mu r)\,.
\end{equation}
In the main text, we fixed the scale parameter $\mu=1$ without lost
of generality. Notice that $\left\| \Phi_+ \right\|= \left\| \Phi_-
\right\|$.

\textit{\textbf{Self-adjoint extensions}}. Hence, the
\emph{deficiency subspaces} $\mathcal{K}_\pm$ are one-dimensional
for $\alpha\in(0,1)|_{\mbox{mod}\ 2}$. We remind that $Q$ coincides
formally with $\tilde{Q}_{\alpha}$ (defined in (\ref{Qgamma})) for
this value of the magnetic flux. The \emph{deficiency indices} are
equal to one, $n_\pm := {\rm dim}\, \mathcal{K}_\pm = 1$, and,
according to von Neumann's theory of self-adjoint extensions of
symmetric operators \cite{R-S}, the self-adjoint extensions of
$\tilde{Q}_{\alpha}$ are characterized by the isometries
$\mathcal{K}_+ \rightarrow \mathcal{K}_-$ (which, in the present
case, form a group $U(1)$ whose elements correspond to a phase
factor $e^{i \gamma}$). Let us denote these self-adjoint
extensions by $Q_{\alpha}^{\gamma}$. Their domain of definition has
the following form

\begin{equation}\label{4-1}
    \begin{array}{c}
      \displaystyle
      \mathcal{D}\left( Q^\gamma_{\alpha} \right):=
      \left\{ \Phi(r,\varphi)=f(r,\varphi)+ A \left[\Phi_+(r,\varphi)
    + e^{i \gamma} \Phi_-(r,\varphi) \right]:
    \right.
      \\ \\
      \displaystyle
      \left. f(r,\varphi)\in \mathcal{D}\left( \overline{Q} \right)\,,
      A\in \mathbb{C} \,, \gamma\in [0,2\pi) \right\}\,.
    \end{array}
\end{equation}
The domain of definition of $\overline{Q}$ for these values of the
flux is given by (\ref{2-9}), (\ref{a-3}), (\ref{a-4}) and
(\ref{9-12}). The operator $Q^{\gamma}_{\alpha}$ acts as
\begin{equation}\label{4-2}
    Q^\gamma_{\alpha} \Phi(r,\varphi):= Q^\dagger \Phi(r,\varphi)=
    \overline{Q} f(r,\varphi)
    + i \mu A \left[\Phi_+(r,\varphi)
    - e^{i \gamma} \Phi_-(r,\varphi) \right]\,.
\end{equation}
Taking into account (\ref{3-8}), the domain can be written as
\begin{eqnarray}\label{domm}
    \mathcal{D}\left(Q_{\alpha}^{\gamma}\right)
    &=&\left\{f(r,\varphi)+A\left(K_{2l_0+\alpha}e^{2l_0i\varphi}
    (1+e^{i\gamma})+K_{1-2l_0-\alpha}e^{(2l_0-1)i\varphi}(1-e^{i\gamma})
    \right)\right\}\,,
\end{eqnarray}
where $f(r,\varphi)$ is from $\mathcal{D}(\overline{Q})$.

\setcounter{equation}{0}
\renewcommand{\theequation}{B.\arabic{equation}}

\section*{Appendix B}
Let us take $\gamma$ as a free parameter. Hamiltonian
$H_{\alpha}^{\gamma}$ is self-adjoint as it is a square of
self-adjoint supercharge  $\tilde{Q}_{\alpha}^{\gamma}$. We can
define the domain $\mathcal{D}_{c}$, see (\ref{DC}), for the current
extension $H_{\alpha}^{\gamma}$. It is dense in $L_2(R^2)$ as it
contains infinitely smooth functions with compact support  as well.

However, $\mathcal{D}_{c}$ cannot accomodate the wave packets
(normalizable combinations of scattering states) for general value
of $\gamma$. Let us demonstrate this
in the following way: we
restrict $\alpha\in[0,1)$. Let
$\Phi_{0}(r,\varphi)=\phi_{0}(r)+\phi_{-1}e^{-i\varphi}$ be a
function lying in the intersection of $\mathcal{H}_0$,
$\mathcal{D}(\tilde{Q}^{\gamma}_{\alpha})$ and $\overline{D}$. It
has the asymptotic behavior at the origin prescribed by (\ref{4-4}).
Acting with $\overline{D}$ we get
\begin{equation}
     \overline{D}\phi_0(r)\sim A(1-\alpha)(1+e^{i\gamma})
     \frac{\Gamma(\alpha)}{2^{1-\alpha}}r^{-\alpha}(1+O(r^2))\,,
\end{equation}
\begin{equation}
     \overline{D}\phi_{-1}(r)\sim A\alpha(1-e^{i\gamma})
     \frac{\Gamma(1-\alpha)}{2^{\alpha}}r^{-1+\alpha}(1+O(r^2))\,.
\end{equation}
\noindent We require that the resulting function does not leave the
domain of definition of $Q_{\alpha}^{\gamma}$. It is a necessary
condition to keep the wave packets composed of scattering states
from $\mathcal{H}_0$ within $\mathcal{D}_{c}$. Considering $\alpha$
as a free parameter, this requirement can be satisfied just for
$\gamma=0$ or $\gamma=\pi$. There exists another possibility as
well: when $\alpha=1/2$, $\overline{D}\Phi_0$ satisfies (\ref{4-4})
for any value of $\gamma$. This  is in agreement with our
observation of  Section \ref{semi-integer}, where the broader family
of the self-adjoint extensions with hidden superysymmetry generator
$\mathcal{U}\tilde{Q}^{0}_{1/2}\mathcal{U}^{-1}$ was revealed in the
case of the half-integer flux. In particular, we discussed the
systems associated with $\tilde{Q}^{\pi/2}_{1/2}$ and
$\tilde{Q}^{3\pi/2}_{1/2}$.
\vspace*{4mm}

Let us  present here the domains of definitions of the operators
$D$, $S_1$, $F$ and ${\cal Z}$\,:
  \begin{equation}
   \mathcal{D}(\overline{D})=\mathcal{D}(D^{\dagger})=
   \{\psi(r,\varphi)\in \mathcal{AC}(\mathbb{R}^2\setminus\{0\})
   \cap L_2(\mathbb{R}^2)|r\partial_r\psi(r,\varphi)\in
   L_2(\mathbb{R}^2)\}\,,
  \end{equation}
\begin{equation}
     \mathcal{D}(S_2)=\mathcal{D}(S_1)=\{\Psi\in L_2(R^2):
     S_1\Psi\in L_2(R^2)\}\,,
\end{equation}
\begin{equation}
     \mathcal{D}(F)=\{\Psi\in L_2(R^2):
     F\Psi\in L_2(R^2)\}\,,
\end{equation}
\begin{equation}
     \mathcal{D}({\cal Z})=\{\Psi\in L_2(R^2):
     {\cal Z}\Psi\in L_2(R^2)\}\,.
\end{equation}

\end{document}